\documentstyle[epsfig]{l-aa}
\let\ov=\over
\let\l=\left
\let\r=\right

\def\be{\begin{equation}}
\def\ee{\end{equation}}
\def \w#1{{\bf #1}}
\def\balpha{{\bar \alpha}}
\def\bdelta{{\bar \delta}}
\def\spose#1{\hbox to 0pt{#1\hss}}
\def\lta{\mathrel{\spose{\lower 3pt\hbox{$\mathchar"218$}}
     \raise 2.0pt\hbox{$\mathchar"13C$}}}
\def\gta{\mathrel{\spose{\lower 3pt\hbox{$\mathchar"218$}}
     \raise 2.0pt\hbox{$\mathchar"13E$}}}

\begin{document}

\thesaurus{06(02.07.1; 02.13.1; 02.18.5; 08.14.1; 08.16.6)}

\title{Gravitational waves from pulsars: emission by the 
magnetic field induced distortion} 

\author{S.~Bonazzola \and E.~Gourgoulhon\thanks{author to whom the proofs 
should be sent} }

\offprints{E.~Gourgoulhon}

\institute{D\'epartement d'Astrophysique Relativiste et de Cosmologie
  (UPR 176 du C.N.R.S.), Observatoire de Paris, \\
   Section de Meudon, F-92195 Meudon Cedex, France \\
   {\em e-mail : bona,gourgoulhon@obspm.fr} }

\date{Received 6 November 1995 / Accepted 20 February 1996}

\maketitle
\markboth{S.~Bonazzola \& E.~Gourgoulhon: Gravitational waves from
pulsars}{S.~Bonazzola \& E.~Gourgoulhon: Gravitational waves from pulsars}

\begin{abstract}
The gravitational wave emission by a distorted rotating fluid star is computed. 
The distortion is supposed to be symmetric around some axis inclined 
with respect to the rotation axis. In the general case, 
the gravitational radiation is emitted at
two frequencies: $\Omega$ and $2\Omega$, where $\Omega$ is the rotation 
frequency. The obtained formul\ae\  are applied to the specific case of a
neutron star distorted by its own magnetic field. Assuming that the
period derivative $\dot P$ of pulsars is a measure of their 
magnetic dipole moment, 
the gravitational wave amplitude can be related to the observable
parameters $P$  and $\dot P$ and to a factor $\beta$
which measures the efficiency of a given magnetic
dipole moment in distorting the star. $\beta$ depends on the nuclear matter 
equation of state and on the magnetic field distribution. 
The amplitude at the frequency $2\Omega$,
expressed in terms of $P$, $\dot P$ and $\beta$, is independent of the
angle $\alpha$ between the magnetic axis and the rotation axis, whereas
at the frequency $\Omega$, the amplitude increases as $\alpha$ decreases. 
The value of $\beta$ for specific models of magnetic field distributions 
has been computed by means of a numerical code giving self-consistent
models of magnetized neutron stars within general relativity. 
It is found that the distortion at fixed magnetic dipole moment
is very dependent of the magnetic field 
distribution; a stochastic magnetic field or
a superconductor stellar interior greatly increases $\beta$ with respect to the 
uniformly magnetized perfect conductor case and
might lead to gravitational waves detectable
by the VIRGO or LIGO interferometers. The amplitude modulation of the signal 
induced by the daily rotation of the Earth has been computed and specified to 
the case of the Crab pulsar and VIRGO detector.
 
\end{abstract}

\section{Introduction}

Rapidly rotating neutron stars (pulsars) might be an important source of
continuous gravitational waves in the frequency bandwidth of
the forthcoming LIGO and VIRGO
interferometric detectors (cf. Bonazzola \& Marck 1994 
for a recent review about the astrophysical sources
these detectors may observe). 
It is well known that a stationary
rotating body, perfectly symmetric with respect to its rotation axis
does not emit any gravitational wave. Thus in order to radiate gravitationally
a pulsar must deviate from axisymmetry. 
Various kinds of pulsar asymmetries have been suggested in the literature:
first the crust of a neutron star is solid, so that its shape may not 
be necessarily axisymmetric under the effect of rotation, as it would
be for a fluid: deviations from axisymmetry are supported by anisotropic
stresses in the solid. The shape of the crust depends not only on the 
geological history of the neutron star, especially on the episode of 
crystallization
of the crust, but also on star quakes. 
Due to its violent formation (supernova) or due to its environment
(accretion disk), the
rotation axis may not coincide with a principal axis of the neutron
star moment of inertia and the star may precess (Pines \& Shaham 1974). 
Even if it keeps a perfectly axisymmetric shape, a freely precessing body
radiates gravitational waves (Zimmermann \& Szedenits 1979).
Neutron stars are known to have important magnetic fields; 
magnetic pressure (Lorentz forces exerted on the conducting
matter) can distort the star 
if the magnetic axis is not aligned with the rotation axis, which is 
widely supposed to occur in order to explain the pulsar phenomenon.
Another mechanism for producing asymmetries is the development of 
non-axisymmetric instabilities in rapidly rotating neutron stars driven
by the gravitational radiation reaction (CFS instablity, cf. e.g.
Schutz 1987) or by nuclear
matter viscosity (cf. e.g. Bonazzola, Frieben \& Gourgoulhon 1996
and references therein).

In the seventies it was widely thought that the solid part of neutron stars
was large (see Sect.~III of Goldreich 1970). A rotating rigid body whose
rotation axis does not coincide with a principal axis of the moment
of inertia must have a precessional motion. Therefore, there have been
a number of studies about the precession of neutron stars and the
resulting gravitational wave emission (Zimmermann 1978\footnote{the results
presented in this reference are erroneous and are corrected in
Zimmermann \& Szedenits (1979)},
Zimmermann \& Szedenits 1979,
Zimmermann 1980, Alpar \& Pines 1985, Barone et al. 1988, 
de Araujo et al. 1994).
However, modern dense matter calculations reveal that 
neutron star interiors are completely liquid 
(see e.g. Haensel 1995). 
Since the crust represents only 1 to 2\% of the stellar mass 
(Lorenz et al. 1993),
it appears that most of the neutron star is in a liquid phase.
In this case, the star is not expected to precess, or at least its
precession frequency must be reduced by a factor $10^5$ as compared
with the solid case (Pines \& Shaham 1974). 

The gravitational radiation from a rotating fluid star has not been
studied as much as that from a solid star. The only reference we are aware of
is the work by Gal'tsov, Tsvetkov \& Tsirulev (1984) and by
Gal'tsov \& Tsvetkov (1984) about the gravitational emission of a
rotating Newtonian homegeneous fluid drop with an oblique magnetic
field. These authors have shown that the gravitational waves are
emitted at two frequencies: the rotation frequency and twice the rotation
frequency. They did not give explicit formul\ae\ for the two 
gravitational wave amplitude $h_+$ and $h_\times$ but
have instead computed the response of a heterodyne
detector (consisting in a rotating dumbbell orientated at right angles
to the direction of propagation of the wave) directly from the 
Riemann tensor.

In the present article, we compute the gravitational radiation from 
a rotating fluid star, distorted along a certain direction by some 
process (for example an internal magnetic field), the angle between
this direction and the rotation axis being arbitrary (Sect.~\ref{s:generation}).
We do not
assume that the star is a Newtonian object: it can have a large
gravitational field described by the theory of general relativity --- 
which is relevant for neutron stars. As an illustration we give the
specific example of a Newtonian incompressible fluid with a uniform
internal magnetic field (Sect.~\ref{s:incomp}). 
We consider also the (more general) case
where the deformation is due to an internal magnetic field which is
also responsible for the $\dot P$ of the pulsar
(electromagnetic braking) (Sect.~\ref{s:magnetic}). Using the numerical 
code presented in (Bocquet, Bonazzola, Gourgoulhon \& Novak 1995) for
magnetized rotating neutron stars in general relativity, we compute the
gravitational emission for a specific $1.4\, M_\odot$ neutron star model
and for various configurations of the internal magnetic field. 
Finally, we derive the response of an interferometric detector to the 
gravitational 
signal, taking into account the daily change induced by the Earth rotation 
in the relative orientation of the detector arms 
and the source (Sect.~\ref{s:detect}). 

\section{Generation of gravitational waves by a rotating fluid star}
\label{s:generation} 

\subsection{Thorne's quadrupole moment}

Neutron stars are fully relativistic objects, so that the standard
quadrupole formula for gravitational wave generation --- which is
derived for non relativistic sources (cf. Sect.~36.10 of
Misner, Thorne \& Wheeler 1973, hereafter MTW) --- does not a priori apply.
However Ipser (1971) has shown that the gravitational radiation from 
a slowly rotating and fully relativistic star is given by a formula 
structurally identical to the weak field standard formula, provided that
the involved multipole moments are defined in an appropriate way. 

The leading term in the gravitational radiation field is then given by
the formula\footnote{The following conventions are used:
Latin indices ($i,j,k,l,\ldots$) range from 1 to 3 and a summation is to 
be performed on repeated indices.}:
\be \label{e:form,quad}
	h_{ij}^{\rm TT} = {2 G \ov c^4} \, {1\ov r} \, 
	   \l[ P_i^{\ \, k} P_j^{\ \, l} - {1\ov 2} P_{ij} P^{kl} \r] 
	{\ddot {\cal I}}_{kl} \l( t- {r\ov c} \r) \ ,
\ee
where $r$ is the distance to the source and
$P_{ij} := \delta_{ij} - r_i r_j/r^2$ is the tensor of projection
transverse to the line of sight. Eq.~(\ref{e:form,quad}) holds for 
highly relativistic sources provided that ${\cal I}_{ij}$ is the
mass quadrupole moment defined as the quadrupolar part of the $1/r^3$ 
term of the $1/r$ expansion of the metric coefficient $g_{00}$ in 
an {\em asymptotically Cartesian and mass centered (ACMC)} coordinate system
(Thorne 1980). This latter is a very broad coordinate class, which 
includes the harmonic coordinates. 
The precise definition of ${\cal I}_{ij}$ is the following one.
The space around a neutron star of mass $M$, ``typical'' radius $R$ and 
angular velocity $\Omega$ can be divided in three regions:
\begin{itemize}
\item the {\em strong-field region}: $r \lta \mbox{a few} \ {G M / c^2}$;
\item the {\em weak-field near zone}: 
$\max(R,\ \mbox{a few} \ {G M / c^2}) \lta r \lta c / \Omega$
\item the {\em wave zone} : $r \gta c / \Omega$
\end{itemize}
In the wave zone, retardation effects are important whereas in the
weak-field near zone, the gravitational field can be considered as
quasi-stationary. Note that for the fastest millisecond 
pulsar to date, PSR 1937+21,
$c / \Omega \sim 80\ {\rm km}$, whereas $R\sim 10{\ \rm km}$ and if its mass is 
$M = 1.4 M_\odot$, $GM/c^2 \sim 2 \ {\rm km}$. Hence, even in this 
extreme case, the weak-field near zone is well defined.
A coordinate system $(t,x^1,x^2,x^3)$ is said to be
{\em asymptotically Cartesian and mass centered (ACMC)} to order 1
if the metric admits the following $1/r$ expansion ($ r:= x_i x^i$)
{\em in the weak-field near zone} (Sect.~XI of Thorne 1980):
\begin{eqnarray}
   g_{00} & = & - 1 + {2M\ov r} + {\alpha_1\ov r^2} \nonumber \\
	 & &	+ {1\ov r^3} \l[ 3\, {\cal I}_{ij} {x^i x^j\ov r^2}
		  + \beta_{1i} {x^i\ov r} + \alpha_2 \r]
		  + O\l( {1\ov r^4} \r) \label{e:def:ACMC:g_00} \\
   g_{0i} & = & - 4 \epsilon_{ikl} J^k {x^l\ov r^3} + O\l( {1\ov r^3} \r) \\
   g_{ij} & = & \delta_{ij} 
	+ {\alpha_{3ij} \ov r} 
	+ {1\ov r^2} \l[ \beta_{2ijk} {x^k\ov r}+ \alpha_{4ij} \r] 
	+ O\l( {1\ov r^3} \r) \ , \label{e:def:ACMC:g_ij}
\end{eqnarray}
where $\alpha_1$, $\alpha_2$, $\alpha_{3ij}$, $\alpha_{4ij}$, 
$\beta_{1i}$ and $\beta_{2ijk}$ are some constants. 
The coefficients $M$ and $J^k$ in the
above expansion are the star's total mass and angular momentum. They are
independent of the $(x^i)$ coordinate choice, provided that they reduce
to usual Cartesian coordinates at the asymptotically flat infinity.
On the contrary, the coefficient ${\cal I}_{ij}$ is not invariant under
change of asymptotically Cartesian coordinates: it is invariant only
with respect to the sub-class of ACMC coordinates. This quantity
is called the {\em mass quadrupole moment} of the star. 
As discussed in Sect.~XI of Thorne (1980) (cf. also Thorne \& G\"ursel 1983),
${\cal I}_{ij}$ is a flat-space-type tensor which ``resides'' in the
weak-field near zone, which means that it can be manipulated as a tensor
in flat space. This tensor is symmetric and trace-free.
If the star
has a gravitational field weak enough to be correctly described by
the Newtonian theory of gravitation, ${\cal I}_{ij}$ is expressible
as minus the trace-free part of the moment of inertia tensor $I_{ij}$:
\be \label{e:calIij,Newt}
    {\cal I}_{ij} = - I_{ij} + {1\ov 3} I_k^{\ k} \, \delta_{ij} \ ,
\ee
$I_{ij}$ being given by
\be \label{e:Iij,Newt}
    I_{ij} := \int \rho \l( x_k x^k \, \delta_{ij} - x_i x_j \r) d^3 x \ .
\ee
In particular, ${\cal I}_{ij}$ coincides at the Newtonian limit with
the quantity $-\!\!\!\! I_{ij}$ introduced in MTW. For highly relativistic
configurations, ${\cal I}_{ij}$ can no longer be expressed as some integral
over the star, as in Eqs.~(\ref{e:calIij,Newt})-(\ref{e:Iij,Newt}); it must
be read from the $1/r$ expansion of the metric coefficient $g_{00}$ in an
ACMC coordinate system. In Appendix \ref{s:appendQI}, 
we give the transformation from
a quasi-isotropic coordinate system, usually used in relativistic studies
of rotating neutron stars (cf the discussion in Sect~2 of
Bonazzola, Gourgoulhon, Salgado \& Marck 1993), to an ACMC coordinate 
system. 

\subsection{Slightly deformed rotating star} \label{s:slight}

Very tight limits have been set on the deformation
of neutron stars by pulsar timings: the relative deviation from
axisymmetry is at most $10^{-3}$ (cf. Table~1 of New et al. 1995). 
This number is certainly overestimated by several orders of magnitude for it 
is obtained by assuming that all the observed period derivative $\dot P$ 
is due to the gravitational emission, whereas it is much more 
likely to be due to the electromagnetic emission. Indeed, the present 
measurements of the {\em braking index} $n$ of pulsars are $n=2.509\pm 0.001$
for the Crab, $n=2.01\pm 0.02$ for PSR 0540-69 and
$n=2.837\pm 0.001$ for PSR 1509-58 (cf. Muslimov \& Page 1996 for references). 
Thus they are closer to $n=3$ (magnetic dipole
radiation) than to $n=5$ (gravitational radiation, cf. Table~9.1 of
Manchester \& Taylor 1977). 

For such small deformations, we make
the assumption that 
the quadrupole moment ${\cal I}_{ij}$ can be linearly decomposed
into the sum of two pieces:
\be \label{e:decomp,Iij}
  {\cal I}_{ij} = {\cal I}_{ij}^{\rm rot} + {\cal I}_{ij}^{\rm dist} \ ;
\ee 
${\cal I}_{ij}^{\rm rot}$ is the quadrupole moment due to rotation:
${\cal I}_{ij}^{\rm rot} = 0$ if the configuration is static.
${\cal I}_{ij}^{\rm dist}$ is the quadrupole moment due to some
process that distorts the star, for example an internal magnetic field
or anisotropic stresses for the nuclear interactions.

The fact that the star does not precess implies
that there exists some ACMC coordinate system $x^\alpha=(t,x,y,z)$ such 
that the components ${\cal I}_{ij}^{\rm rot}$ in this coordinate system
do not depend upon $x^0 = t$ and take the diagonal form
\be \label{e:Iijrot}
   {\cal I}_{ij}^{\rm rot} = \l( \begin{array}{ccc} 
	- {\cal I}_{zz}^{\rm rot} / 2 & 0 & 0 \\
	0 & - {\cal I}_{zz}^{\rm rot} / 2 & 0 \\
	0 & 0 & {\cal I}_{zz}^{\rm rot} 
      \end{array} \r)  \ .
\ee
The line $x=y=0$ is the rotation axis. For a small angular velocity $\Omega$,
${\cal I}_{zz}^{\rm rot}$ is a quadratic function of $\Omega$.

\begin{figure}
\epsfig{figure=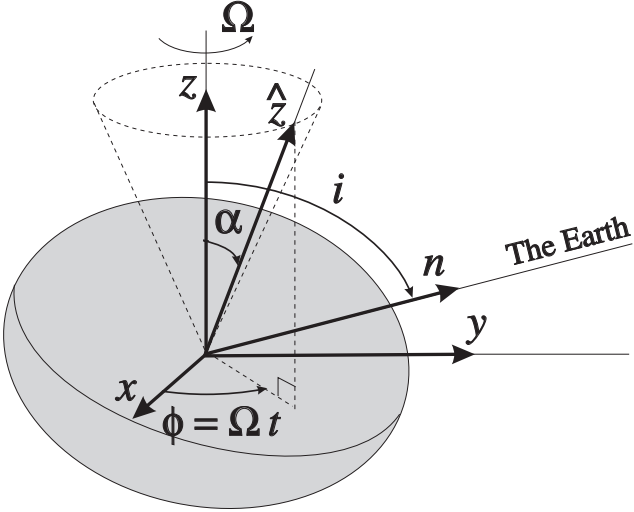,height=7cm}
\caption[]{\label{f:e_hat}
Geometry of the distorted neutron star.}    
\end{figure}

Let us make the following assumptions about the distortion process:
\begin{enumerate}
\item there exists some privileged direction, i.e. two of the three 
eigenvalues of ${\cal I}_{ij}^{\rm dist}$ are equal, i.e. there
exists in the weak-field near zone a coordinate system (not necessarily ACMC)
$x^{\hat \alpha} = (\hat t,\hat x, \hat y,\hat z)$
such that the components ${\cal I}_{\hat i\hat j}^{\rm dist}$ are
\be \label{e:I_hihj}
   {\cal I}_{\hat i\hat j}^{\rm dist} = \l( \begin{array}{ccc} 
	- {\cal I}_{\hat z\hat z}^{\rm dist} / 2 & 0 & 0 \\
	0 & - {\cal I}_{\hat z\hat z}^{\rm dist} / 2 & 0 \\
	0 & 0 & {\cal I}_{\hat z\hat z}^{\rm dist} 
      \end{array} \r)  \ .
\ee
\item the deformation is rotating with the star, i.e. the weak-field
near zone vector $\w{e}_{\hat z}$ associated with the coordinate
$\hat z$ is rotating at the angular velocity
$\Omega$ at a fixed angle $\alpha$ from the weak-field
near zone vector $\w{e}_z$ associated with the coordinate $z$ ($\w{e}_z$
is along the rotation axis) (cf. Fig.~\ref{f:e_hat}):
\be
    \w{e}_{\hat z} = \cos\alpha \, \w{e}_z + \sin\alpha \l[
		\cos\varphi(t)\, \w{e}_x + \sin\varphi(t)\,  \w{e}_y \r] \ ,
\ee
with
\be \label{e:phi(t)}
	\varphi(t) = \Omega (t - t_0) \ .
\ee
\end{enumerate}
In particular, the above assumptions are satisfied by a magnetic field symmetric
with respect to some axis: $\w{e}_{\hat z}$ is then parallel to the
magnetic dipole moment vector {\boldmath $\cal M$}.

\subsection{Application of the quadrupole formula}

Inserting Eq.~(\ref{e:decomp,Iij}) into the quadrupole formula 
(\ref{e:form,quad}) and using the time constancy of ${\cal I}_{ij}^{\rm rot}$,
we get
\be \label{e:quad,dist}
	h_{ij}^{\rm TT} = {2 G \ov c^4} \, {1\ov r} \, 
	   \l[ P_i^{\ \, k} P_j^{\ \, l} - {1\ov 2} P_{ij} P^{kl} \r] 
	{\ddot {\cal I}}_{kl}^{\rm dist} \l( t- {r\ov c} \r) \ .
\ee
In order to apply this formula, one must first express 
${\cal I}_{ij}^{\rm dist}$ in the ACMC coordinates $(t,x,y,z)$, from its
components (\ref{e:I_hihj}) in the coordinates 
$(\hat t, \hat x, \hat y, \hat z)$. The transformation 
$(\hat t, \hat x, \hat y, \hat z)\rightarrow(t,x,y,z)$ is simply the
composition of two rotations (cf. Fig.~\ref{f:e_hat}): one of angle 
$\alpha$ around an axis $\w{e}_{x'}$ in the $(\w{e}_x,\w{e}_y)$
plane and rotating with the star, and one of angle 
$\varphi(t) = \Omega (t-t_0)$ around $\w{e}_{z}$.
The transformation matrix is then:
\be
   P = \l( \begin{array}{ccc} 
	\cos\varphi(t)	& -\sin\varphi(t)	& 0	\\
	\sin\varphi(t)	& \cos\varphi(t)	& 0	\\
	0		& 0		& 1	
       \end{array} \r)
	\times
       \l( \begin{array}{ccc}
	1	& 0 		& 0		\\
	0	& \cos\alpha	& \sin\alpha	\\
	0	& -\sin\alpha	& \cos\alpha	
       \end{array} \r) .
\ee
The tensor transformation law writes 
${\cal I}^{\rm dist} = P \times \hat{\cal I}^{\rm dist}\times {}^t P$, 
where ${\cal I}^{\rm dist}$ is the
matrix formed by the components ${\cal I}_{ij}^{\rm dist}$ and 
$\hat{\cal I}^{\rm dist}$ is
the matrix formed by the components ${\cal I}_{\hat i\hat j}^{\rm dist}$
[Eq.~(\ref{e:I_hihj})]. 
Performing this matrix product leads to
\begin{eqnarray}
  {\cal I}_{ij}^{\rm dist} \!\!\!& & = 
	{1\ov 2} {\cal I}_{\hat z\hat z}^{\rm dist}
	\l( \begin{array}{c}
 3\sin^2\alpha\sin^2\varphi(t) - 1	  	\\
 - 3/2\, \sin^2\alpha\sin2\varphi(t)  	\\
 - 3\sin\alpha\cos\alpha\sin\varphi(t)  
	\end{array} \r. \nonumber \\
  & & \l. \begin{array}{cc}
  -3/2\, \sin^2\alpha\sin2\varphi(t) & - 3\sin\alpha\cos\alpha\sin\varphi(t) \\
  3\sin^2\alpha\cos^2\varphi(t)- 1 & 3\sin\alpha\cos\alpha\cos\varphi(t) \\
  3\sin\alpha\cos\alpha\cos\varphi(t) & 3\cos^2\alpha - 1
	\end{array} \r)	. \label{e:Iijdist} 
\end{eqnarray}
In order to apply the quadrupole formula, the second derivative with 
respect to $t$ of this expression must be taken. One obtains, 
using Eq.~(\ref{e:phi(t)}),
\begin{eqnarray}
  {\ddot {\cal I}}_{ij}^{\rm dist} \!\!\!& & = 
	{3\ov 2} {\cal I}_{\hat z\hat z}^{\rm dist} \, \Omega^2 \sin\alpha
	\l( \begin{array}{c}
	2\sin\alpha\cos2\varphi(t) \\
	2\sin\alpha\sin2\varphi(t) \\
	\cos\alpha\sin\varphi(t)
	\end{array} \r. \nonumber \\
  & & \qquad \qquad \l. \begin{array}{cc}
 2\sin\alpha\sin2\varphi(t) & \cos\alpha\sin\varphi(t)	\\
 -2\sin\alpha\cos2\varphi(t) & -\cos\alpha\cos\varphi(t)	\\
  -\cos\alpha\cos\varphi(t) & 0 		
	\end{array} \r)	. \label{e:ddotIij}
\end{eqnarray}
The next step consists in taking the transverse traceless projection
of ${\ddot {\cal I}}_{ij}^{\rm dist}$. Let us denote by $i$ the angle
between the neutron star's rotation axis $\w{e}_z$ and the direction
$\w{n}$ from
the star's centre to the Earth ($i$ is called the {\em line of
sight inclination}) (cf. Fig.~\ref{f:e_hat}). Without any loss of generality,
we can choose the ACMC coordinates
$(t,x,y,z)$ such that $\w{n}$ lies in the $(\w{e}_y,\w{e}_z)$ plane.
The components with respect to $(t,x,y,z)$ of the transverse projection 
operator $P_{ij} = \delta_{ij} - n_i n_j$ are then
\be \label{e:Pij}
	P_{ij} = \l( \begin{array}{ccc}
    1 	& 0		& 0 		\\
    0 	& \cos^2 i 	& -\sin i\cos i \\
    0   & -\sin i\cos i & \sin^2 i
		\end{array} \r) \ .
\ee
From Eqs.~(\ref{e:Pij}) and (\ref{e:ddotIij}), 
the computation of the right-hand side of the quadrupole formula
(\ref{e:quad,dist}) is straightforward, though somewhat tedious.
The result is
\be 
   h_{ij}^{\rm TT} = h_+ \, e_{ij}^+ \ + \ h_\times \, e_{ij}^\times \ ,
\ee
with
\begin{eqnarray}
   e_{ij}^+ & = & \l( \begin{array}{ccc}
		1	& 0		& 0		\\
		0	& -\cos^2 i	& \sin i\cos i	\\
		0	& \sin i\cos i  & - \sin^2 i	
	\end{array} \r) \ , \\
   e_{ij}^\times & = & \l( \begin{array}{ccc}
	        0	& \cos i	& - \sin i	\\
		\cos i	& 0		& 0		\\
		-\sin i & 0		& 0	
	\end{array} \r) 
\end{eqnarray}
and
\begin{eqnarray}
   h_+ & = & h_0 \sin\alpha \Big[
	{1\ov 2} \cos\alpha\sin i\cos i \cos\Omega(t-t_0) \nonumber \\
	& & \qquad \qquad
 - \sin\alpha {1+\cos^2 i\ov 2} \cos2\Omega (t-t_0) \Big] \label{e:h+,gen} \\
   h_\times & = & h_0 \sin\alpha \Big[
	{1\ov 2} \cos\alpha\sin i\sin\Omega(t-t_0) \nonumber \\
	& & \qquad \qquad
	- \sin\alpha \cos i \sin2\Omega (t-t_0) \Big] \ , \label{e:hx,gen}
\end{eqnarray}
where
\be \label{e:def:h0}
	h_0 := - {6 G \ov c^4} {\cal I}_{\hat z\hat z}^{\rm dist}
		{\Omega^2\ov r} \ .
\ee
Note that in the above expressions,
Eq.~(\ref{e:phi(t)}) has been substituted for $\varphi(t)$ and that the
retardation term $-r/c$ has been incorporated into the constant $t_0$.

\subsection{Discussion}

From the formul\ae\ (\ref{e:h+,gen})-(\ref{e:hx,gen}), it is clear that
there is no gravitational emission if the distortion axis is aligned
with the rotation axis ($\alpha=0\ \mbox{or}\ \pi$). If both axes are
perpendicular ($\alpha=\pi/2$), the gravitational emission is monochromatic
at twice the rotation frequency. In the general case ($0<|\alpha|<\pi/2$),
it contains two frequencies: $\Omega$ and $2\Omega$. For small values
of $\alpha$ the emission at $\Omega$ is dominant.

It may be noticed that Eqs.~(\ref{e:h+,gen})-(\ref{e:def:h0}) are
structurally equivalent to Eq.~(1) of Zimmermann \& Szedenits (1979) 
(hereafter ZS),
although this latter work is based on a different physical hypothesis
(Newtonian precessing rigid star). In order to compare precisely 
Eqs.~(\ref{e:h+,gen})-(\ref{e:def:h0}) and Eq.~(1) of ZS, some slight
re-arrangements must be performed. First, the ellipticity $\epsilon$
defined by ZS is linked to our ${\cal I}_{\hat z\hat z}$ by
$I_1 \epsilon = -3/2\, {\cal I}_{\hat z\hat z}$ 
[cf. Eq.~(\ref{e:calIij,Newt})], this explains the factor 2 in front of
the right-hand side of Eq.~(1) of ZS instead of the factor $-6$ in 
Eq.~(\ref{e:def:h0}). Other apparent differences
are actually due to different conventions: the origin of time in ZS is 
the instant when the equivalent of the deformation axis is at its farthest position
from the observer, whereas in our case it corresponds to its nearest
position. One must then make $t\rightarrow t+\pi$ in order to compare
the two formul\ae. Moreover the choice for the matrices $e_{ij}^+$ and
$e_{ij}^\times$ are exactly opposite in both approaches [cf. Sect.~III.B of
Zimmermann (1980)], so that the transforms $h_+ \rightarrow -h_+$ and
$h_\times \rightarrow -h_\times$ must also be performed. When all this
is done, Eqs.~(\ref{e:h+,gen})-(\ref{e:def:h0}) and Eq.~(1) of ZS appear
to have exactly the same structure. However, the physical significance is 
different: the frequency $\omega$ which appears in Eq.~(1) of ZS differs 
from the pulsar frequency by the body-frame precessional frequency 
whereas in Eqs.~(\ref{e:h+,gen})-(\ref{e:def:h0}), $\Omega$ is exactly
the pulsar frequency. The angle $\theta$ of ZS, which in their Eq.~(1)
takes the place of our angle $\alpha$ in 
Eqs.~(\ref{e:h+,gen})-(\ref{e:def:h0}), is the angle between the total
angular momentum $\w{J}$ and the star's third principal axis, whereas
our $\alpha$ is the angle between the rotation axis and the direction of
the distortion, which even in the Newtonian case,
does not coincide with any of the principal axis of the body (except in the
non-rotating case).

As special cases of Eqs.~(\ref{e:h+,gen})-(\ref{e:def:h0}), one may 
recover results previously published in the literature. For instance,
the case of a triaxial star rotating about a principal axis of its
moment of inertia tensor can be obtained by setting $\alpha = \pi/2$
in Eqs.~(\ref{e:h+,gen})-(\ref{e:def:h0}). The result can be compared
with Eqs.~(48) and (54) of Thorne (1987), noticing that $f$ in these
equations is $\Omega/\pi$ and that 
${\cal I}_{\bar x\bar x}$ and ${\cal I}_{\bar y\bar y}$ of Thorne (1987)
are related to our ${\cal I}_{\hat z\hat z}$ by
${\cal I}_{\bar x\bar x} = - {\cal I}_{\hat z\hat z} / 2$
and ${\cal I}_{\bar y\bar y} = {\cal I}_{\hat z\hat z}$. The two formul\ae\
appear then to be identical, as expected. If in addition to $\alpha = \pi/2$,
the inclination angle $i$ is set to zero, 
Eqs.~(\ref{e:h+,gen})-(\ref{e:def:h0}) reduce to the formula used in the 
recent work by New et al. (1995) [their Eq.~(5)]. Note that in the
two studies mentionned above, the gravitational waves are emitted at the
frequency $2\Omega$ only, due to $\alpha=\pi/2$ or equivalently due to the
fact that the rotation axis coincides with a principal axis of the moment
of inertia tensor. Let us stress again that in our (more general) case, the
gravitational radiation contains two frequencies: $\Omega$ and $2\Omega$.

\subsection{Numerical estimates} \label{s:numer,estim}

In order to describe the star deformation by a dimensionless quantity,
let us introduce instead of ${\cal I}_{\hat z\hat z}^{\rm dist}$ the
{\em ellipticity}
\be\label{e:def:eps}
	\epsilon := - {3\ov 2} \, {{\cal I}_{\hat z\hat z}^{\rm dist}
			\ov I}	\ ,
\ee
where $I$ is the moment of inertia with respect to the rotation axis,
defined as
\be \label{e:def:I}
	I := J / \Omega		\ ,
\ee
$J$ being the star angular momentum. The definition (\ref{e:def:I}) is
valid even in highly relativistic cases, provided that the star distortion
is small --- as we suppose throughout this work. Indeed, in this case
the configuration is essentially axisymmetric and the angular momentum
$J$ is well defined (cf. e.g. the discussion in Wald (1984),
p. 297). The factor $-3/2$ in Eq.~(\ref{e:def:eps}) is introduced in order
to recover the classical definition of the ellipticity at the Newtonian
limit (see e.g. Shapiro \& Teukolsky 1983).

Let us introduce also the rotation period $P={2\pi/\Omega}$, since 
observational data about pulsars are usually presented
with $P$ instead of $\Omega$. 

Inserting Eq.~(\ref{e:def:eps}) in expression (\ref{e:def:h0}) for the
characteristic gravitational wave amplitude leads to
\be \label{e:h0,eps}
    h_0 = {16\pi^2 G\ov c^4} {I\, \epsilon\ov P^2\, r} \ .
\ee
Replacing the physical
constants by their numerical values results in  
\be \label{e:h0,num}
    h_0 = 4.21\times 10^{-24} \ \Big[ {{\rm ms}\ov P} \Big] ^2
	\Big[ {{\rm kpc}\ov r} \Big] 
	\Big[ {I\ov 10^{38} {\ \rm kg\, m}^2} \Big]
	\Big[ {\epsilon \ov 10^{-6} } \Big] .
\ee
Note that $I=10^{38} {\ \rm kg\, m}^2$ is a representative value for
the moment of inertia of a $1.4 \, M_\odot$ neutron star [see Fig.~12
of Arnett \& Bowers (1977)].
	
For the Crab pulsar, $P=33 {\ \rm ms}$ and $r=2{\ \rm kpc}$, so that
Eq.~(\ref{e:h0,num}) becomes
\be
   h_0^{\rm Crab} = 1.89\times 10^{-27}  
	\Big[ {I\ov 10^{38} {\ \rm kg\, m}^2} \Big]
	\Big[ {\epsilon \ov 10^{-6} } \Big] .
\ee
For the Vela pulsar, $P=89 {\ \rm ms}$ and $r=0.5{\ \rm kpc}$, hence
\be
   h_0^{\rm Vela} = 1.06\times 10^{-27}  
	\Big[ {I\ov 10^{38} {\ \rm kg\, m}^2} \Big]
	\Big[ {\epsilon \ov 10^{-6} } \Big] .
\ee
For the millisecond pulsar\footnote{We do not consider the
``historical'' millisecond pulsar PSR 1937+21 for it is more than twice
farther away.} PSR 1957+20, $P=1.61 {\ \rm ms}$ and $r=1.5{\ \rm kpc}$, hence
\be \label{e:h0,1957+20}
   h_0^{\mbox{\tiny 1957+20}} = 1.08\times 10^{-24}  
	\Big[ {I\ov 10^{38} {\ \rm kg\, m}^2} \Big]
	\Big[ {\epsilon \ov 10^{-6} } \Big] .
\ee
At first glance, PSR 1957+20 seems to be a much more favorable candidate than
the Crab or Vela. However, in the above formula, $\epsilon$ is in units of 
$10^{-6}$ and the very low value of the period derivative $\dot P$ of
PSR 1957+20 implies that its $\epsilon$ is at most $1.6\times 10^{-9}$
(New et al. 1995). Hence the maximum amplitude one can expect for this
pulsar is $h_0^{\mbox{\tiny 1957+20}} \sim 1.7 \times 10^{-27}$ and
not $1.08\times 10^{-24}$ as Eq.~(\ref{e:h0,1957+20}) might suggest.

\section{The incompressible magnetized fluid example} \label{s:incomp}

As an illustration, let us consider the specific example taken
by Gal'tsov, Tsvetkov \& Tsirulev (1984) (hereafter GTT) and Gal'tsov 
\& Tsevtkov (1984). In their work, a pulsar is idealized as a rigidly
rotating Newtonian body made of an incompressible fluid and endoved with
a magnetic field which is uniform 
inside the star and dipolar oustide it, the magnetic dipole moment 
being inclined by an angle $\alpha$ with respect to the rotation axis.

The rotation rate is supposed to be far from the mass-shedding limit
so that the departure from spherical symmetry is small.
Moreover, the magnetic energy is assumed to be much lower than the 
rotational kinetic energy, which is satisfied by realistic  
configurations. Under these hypotheses, the star takes the shape
of a (quasi-spherical) triaxial ellipsoid. 
The gravitational potential can be then
calculated from the analytical formul\ae\ of Chandrasekhar (1969).
The shape of the ellipsoid is deduced from the first integral of the 
equation of motion
(including the magnetic pressure)
and the magnetic field matching conditions at the stellar surface.
It is found that the ellipsoid is determined by the equation
$(\delta_{ij} + a_{ij}) X^i X^j = R^2$, where (i) $R$ is the
mean radius of the (quasi-spherical) ellipsoid, (ii)
$X^i$ are Cartesian 
coordinates in a co-moving frame, i.e. rotating at the angular velocity
$\Omega$ with respect to an inertial frame and (iii) the $a_{ij}$ are
given by\footnote{Note that the value of $a_{ij}$ presented in GTT [their
Eq.~(2.15)] is erroneous: the $\Omega_i$ part has the wrong sign and the
${\cal M}_i$ part should contain a $R^{-8}$ factor instead of the $R^4$.
This latter error has been corrected in Gal'tsov \& Tsvetkov (1984), but not
the former one.}
\be \label{e:a_ij}
    a_{ij} = {15\ov 2} \, {\Omega_i \Omega_j \ov \omega_{\rm J}^2 }
	+ {45 \mu_0 \ov 32 \pi^2} \, 
        {{\cal M}_i {\cal M}_j \ov R^8 \rho \omega_{\rm J}^2 } \ .
\ee
In this expression, $\rho$ is the constant mass density of the star,
$\omega_{\rm J}=\sqrt{4\pi G\rho}$ is the Jeans frequency, $\Omega_i$ and
${\cal M}_i$ are respectively 
the components of the angular velocity vector and the components
of the magnetic dipole moment with respect to the $X^i$
coordinates: $\Omega_i = (0,0,\Omega)$ and 
${\cal M}_i = (0, {\cal M}\sin\alpha, {\cal M}\cos\alpha)$.
By diagonalizing the matrix $a_{ij}$ given by Eq.~(\ref{e:a_ij}),
one obtains the principal axes of the ellipsoid and the values
of the three semi-axes, $a_1$, $a_2$ and $a_3$. From these quantities
the moment of
inertia tensor $I_{ij}$ can be computed by evaluating the integral
(\ref{e:Iij,Newt}) in the frame of the principal axes. The quadrupole
moment ${\cal I}_{ij}$ is then immediately deduced via 
Eq.~(\ref{e:calIij,Newt}). Transforming the result in the inertial
frame leads to the form (\ref{e:decomp,Iij}) of ${\cal I}_{ij}$,
incidently demonstrating this formula in the particular case under 
consideration,
with the form (\ref{e:Iijrot}) for ${\cal I}_{ij}^{\rm rot}$ with
\be
	{\cal I}_{zz}^{\rm rot} = - {R^5 \Omega^2 \ov 3 G} 
\ee
and with the form (\ref{e:Iijdist}) for ${\cal I}_{ij}^{\rm dist}$ with
\be \label{e:Izz,inc,mag}
	{\cal I}_{\hat z\hat z}^{\rm dist} = - 
	{\mu_0 {\cal M}^2 \ov 16 \pi^2 G \rho R^3} \ .
\ee
The moment of inertia with respect to the rotation axis of the 
homogeneous star is $I = 8\pi \rho R^5 / 15$, so that Eq.~(\ref{e:def:eps})
gives the ellipticity:
\be
   \epsilon = {45\ov 64\pi} {B_{\rm pole}^2 \ov \mu_0 G \rho^2 R^2} \ .
\ee
Note that in this formula, the magnetic dipole moment $\cal M$ has been
expressed in terms of the North pole magnetic field
$B_{\rm pole} = \mu_0/(4\pi)\, 2 {\cal M} / R^3$.

For numerical estimates, let us take typical values for neutron stars:
$B_{\rm pole}=10^9$ T, $M=1.4\, M_\odot$ and
$R=10$ km. Then $\epsilon \simeq 6.0\times 10^{-11}$. This is a very
tiny value, which leads to a gravitational wave amplitude of
only $h_0 \sim 3\times 10^{-30}$ for $P=10$ ms and $r=1$ kpc 
[cf. Eq.~(\ref{e:h0,num})]. However the above model is a very simplified one.
It can be expected that relaxing the assumptions of 
(i) incompressible fluid, (ii) Newtonian gravity and (iii) uniform internal
magnetic field, may lead to a greater value of $\epsilon$.

\section{Magnetic field induced deformation} \label{s:magnetic}

\subsection{Emission formula}

The situation considered in the preceding section is a very simplified one. 
However, one may consider that the obtained form of the deformation, 
Eq.~(\ref{e:Izz,inc,mag}), is qualitatively the same for realistic
magnetized neutron star models, i.e. a compressible perfect fluid
obeying a ``sophisticated'' equation of state resulting from nuclear
physics calculations, and involving general relativity.  More precisely, we 
consider that the magnetic field
induced deformation is a quadratic function of the amplitude of the
magnetic dipole moment, $\cal M$, as in Eq.~(\ref{e:Izz,inc,mag}):
\be
	\epsilon = \beta \, {{\cal M}^2 \ov {\cal M}_0^2} \ ,
\ee
where ${\cal M}_0$ has the dimension of a magnetic dipole moment in order
to make the coefficient $\beta$ dimensionless. Let us choose
${\cal M}_0^2 = (4\pi/\mu_0) \, G I^2 / R^2$ where $R$ is the circumferential
equatorial radius of the star\footnote{For Newtonian stars, $R$ is simply
the equatorial radius; for relativistic stars, $R$ is the length of the
equator, as measured by a locally non rotating observer, divided by $2\pi$.}.
Hence
\be \label{e:eps(M)}
    \epsilon = \beta \, {\mu_0 \ov 4\pi} {{\cal M}^2 R^2\ov G \, I^2} \ .
\ee
Provided the magnetic field amplitude does not take (unrealistic) huge values
($> 10^{14}$ T), this formula is certainly true, even if 
the magnetic field
structure is quite complicated, depending upon the assumed
electromagnetic properties of the fluid: normal conductor, superconductor,
ferromagnetic... The coefficient $\beta$ measures the efficiency of this
magnetic structure in distorting the star. In the following, we shall call 
$\beta$ the {\em magnetic distortion factor}. 
For the simplified model
considered in Sect.~\ref{s:incomp} (incompressible fluid, uniform
internal magnetic field), $\beta=1/5$. 

As argued in Sect.~\ref{s:slight}, 
the observed spin down of radio pulsars is very certainly due
to the low frequency magnetic dipole radiation. $\cal M$ is then linked to the
observed pulsar period $P$ and period derivative $\dot P$ by
[cf e.g. Eq.~(6.10.26) of Straumann (1984)]
\be \label{e:M(Ppoint)}
    {\cal M}^2 = {4\pi \ov \mu_0} {3 c^3\ov 8 \pi^2} 
		{I P \dot P\ov \sin^2\alpha} \ ,
\ee
where $\alpha$ is the angle between the magnetic dipole moment 
{\boldmath $\cal M$} and the
rotation axis. For highly relativistic configurations, the vector 
{\boldmath $\cal M$} is defined in the weak-field near zone (cf. Sect.~2.5
of Bocquet et al. 1995), so is $\alpha$.
Inserting Eqs.~(\ref{e:eps(M)}) and (\ref{e:M(Ppoint)}) 
into  Eq.~(\ref{e:h0,eps}) leads to the gravitational wave amplitude
\be \label{e:th0,def}
  h_0 = 6\beta \, {R^2 \dot P\ov c r P \sin^2\alpha} \ ,
\ee
so that formul\ae\ (\ref{e:h+,gen}) and (\ref{e:hx,gen}) become
\begin{eqnarray}
   h_+ & = & 6\beta \, {R^2 \dot P\ov c r P} \Big[
	{\sin i\cos i \ov 2\tan\alpha} \cos\Omega(t-t_0) \nonumber \\
	& & \qquad \qquad \quad
 -  {1+\cos^2 i\ov 2} \cos2\Omega (t-t_0) \Big] \label{e:h+,mag} \\
   h_\times & = & 6\beta \, {R^2 \dot P\ov c r P} \Big[
	{\sin i\ov 2\tan\alpha} \sin\Omega(t-t_0) \nonumber \\
	& & \qquad \qquad \quad
	-  \cos i \sin2\Omega (t-t_0) \Big] \ . \label{e:hx,mag}
\end{eqnarray}
Equation (\ref{e:th0,def}) can be cast in a numerically convenient form:
\be \label{e:th0,num}
    h_0 = 6.48\times 10^{-30} \, {\beta\ov \sin^2\alpha} \, 
	\Big[ {R\ov 10 {\, \rm km}} \Big] ^2
	\Big[ {{\rm kpc}\ov r} \Big] 
	\Big[ {{\rm ms}\ov P} \Big]
	\Big[ {\dot P\ov 10^{-13}} \Big] \ .
\ee
As a check of this equation, let us consider again the simplified case
of Sect.~\ref{s:incomp}. The uniformly magnetized 
homogeneous Newtonian star with $M=1.4\, M_\odot$,
$R=10\ {\rm km}$, $B_{\rm pole} = 10^9\ {\rm T}$ and $P=10\ {\rm ms}$
has a moment of inertia $I=1.1\times 10^{38}\ {\rm kg\, m}^2$, a factor
$\beta=1/5$, and 
magnetic dipole moment ${\cal M} = 5.0\times 10^{27}\ {\rm A\, m}^2$. 
From Eq.~(\ref{e:M(Ppoint)}), the corresponding period derivative is 
$\dot P = 2.2\times 10^{-12}\, \sin^2\alpha$. Equation~(\ref{e:th0,num}) gives
then $h_0 = 3\times 10^{-30}$ for $r=1\ {\rm kpc}$,
in agreement with the result obtained in Sect.~\ref{s:incomp}. 

The main feature of the emission formul\ae\ (\ref{e:h+,mag})-(\ref{e:hx,mag})
is that the amplitude of the gravitational radiation at the frequency
$2\Omega$ does not depend upon the (unknown) angle $\alpha$. 

\subsection{Discussion}

Among the 706 pulsars of the catalog by Taylor et al. (1995, 1993), the
highest value of $h_0$ at fixed $\alpha$, $\beta$ and $R$, as given by 
Eq.~(\ref{e:th0,num}), is achieved by the Crab pulsar
($P=33 {\ \rm ms}$, $\dot P = 4.21\times 10^{-13}$, 
$r=2{\ \rm kpc}$), followed by Vela ($P=89 {\ \rm ms}$,
$\dot P = 1.25\times 10^{-13}$, $r=0.5{\ \rm kpc}$)
and PSR~1509-58 ($P=151 {\ \rm ms}$,
$\dot P = 1.54\times 10^{-12}$, $r=4.4{\ \rm kpc}$):
\begin{eqnarray} 
   h_0^{\rm Crab} & = & 4.08\times 10^{-31}   
	\Big[ {R\ov 10 {\, \rm km}} \Big] ^2 \, {\beta\ov \sin^2\alpha}
					 \label{e:th0,Crab}  \\
   h_0^{\rm Vela} & = & 1.81\times 10^{-31}  
	\Big[ {R\ov 10 {\, \rm km}} \Big] ^2 \, {\beta\ov \sin^2\alpha}
					 \label{e:th0,Vela}\\
   h_0^{\mbox{\tiny 1509-58}} & = & 1.50\times 10^{-31}  
	\Big[ {R\ov 10 {\, \rm km}} \Big] ^2 \, {\beta\ov \sin^2\alpha} \\
   h_0^{\mbox{\tiny 1957+20}} & = & 4.51\times 10^{-37}  
	\Big[ {R\ov 10 {\, \rm km}} \Big] ^2 \, {\beta\ov \sin^2\alpha} \ .
\end{eqnarray}
We have added to the list the millisecond pulsar
PSR 1957+20 ($P=1.61 {\ \rm ms}$,
$\dot P = 1.68\times 10^{-20}$, $r=1.5{\ \rm kpc}$) considered in 
Sect.~\ref{s:numer,estim}. 
From the above values, it appears that 
PSR 1957+20 is not a good candidate. This is not
suprising since it has a small magnetic field (yielding a low $\dot P$). 
Even for the Crab and Vela pulsars, which have a large $\dot P$, 
the $h_0$ values as given by Eqs.~(\ref{e:th0,Crab}), 
(\ref{e:th0,Vela}) are, at first glance, not very encouraging. Let us
recall that with the $10^{-22} \ {\rm Hz}^{-1/2}$ expected sensitivity of 
the VIRGO experiment at the 30 Hz frequency
(Bondu 1996; see also Fig.~9 of Bonazzola \& Marck 1994), the minimal
amplitude detectable within three years of integration is 
\be \label{e:hmin}
	h_{\rm min} \sim 10^{-26} \ . 
\ee
Comparing this number with Eqs.~(\ref{e:h+,mag})-(\ref{e:hx,mag}) and
(\ref{e:th0,Crab})-(\ref{e:th0,Vela}), one realizes that in order to 
lead to a detectable signal, the angle $\alpha$ must be small and/or
the distortion factor $\beta$ must be large. In the former case, the emission 
is mainly at the frequency $\Omega$. 
From Eqs.~(\ref{e:h+,mag})-(\ref{e:hx,mag}),
the gravitational wave amplitude can even be arbitrary large if 
$\alpha\rightarrow 0$. However, if $\alpha$ is too small, let say 
$\alpha < 10^{-2}$, the simple magnetic braking formula (\ref{e:M(Ppoint)})
certainly breaks down. So one cannot rely on a tiny $\alpha$ to yield a 
detectable amplitude\footnote{From the observed pulse profile and 
polarization of pulsars, values of $\alpha$ can be inferred; they spread
all the range between $0$ and $\pi/2$ (Lyne \& Manchester 1988, Rankin 1990)}.
The alternative solution is to have a large $\beta$. 
Let us recall that for an incompressible fluid with a uniform magnetic field, 
$\beta=1/5$ (Sect.~\ref{s:incomp}). In the following section, we give the 
$\beta$ coefficients computed for
more realistic models (compressible fluid, realistic equation
of state, general relativity taken into account) with various distribution
of the magnetic field. 

\begin{figure}
\epsfig{figure=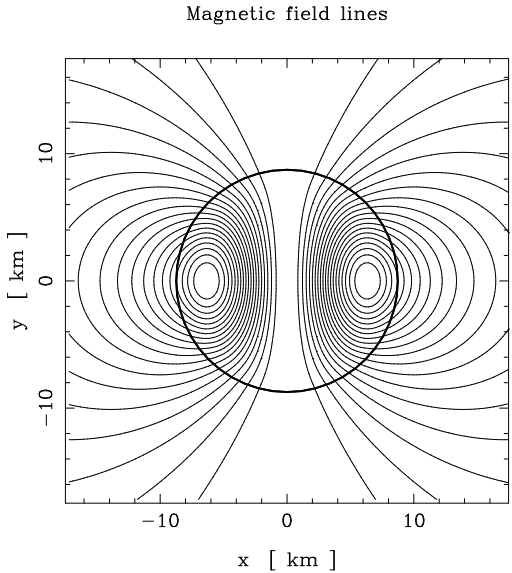,angle=270,height=8cm}
\caption[]{\label{f:mag:f=const}
Magnetic field lines generated by the current distribution corresponding to
the choice $f={\rm const}$. The thick line denotes the star's surface. 
The distortion factor of this configuration is $\beta=1.01$.}
\end{figure}

\begin{figure}
\epsfig{figure=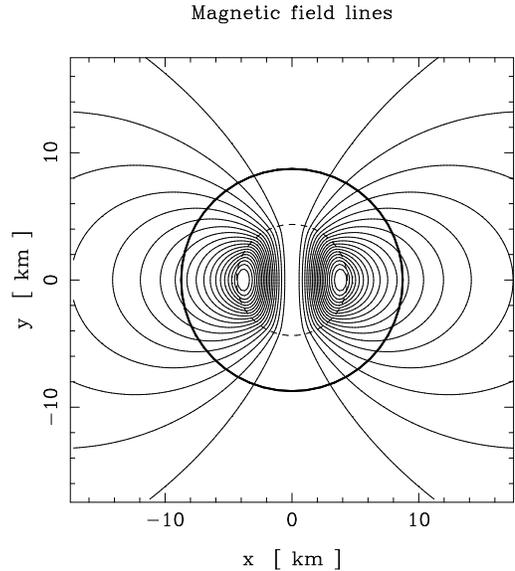,angle=270,height=8cm}
\caption[]{\label{f:mag:j:coeur}
Magnetic field lines generated by a current distribution localized in 
the core of the star and corresponding to
the choice $f={\rm const}$. The thick line denotes the star's surface
and the dashed line the external limit of the electric current distribution.  
The distortion factor corresponding to this situation is $\beta=0.70$.}
\end{figure}

\subsection{Numerical results}

We have developed a numerical code to compute the deformation of 
magnetized neutron stars within general relativity. This code is an
extension of that presented in Bocquet et al. (1995) (hereafter BBGN). 
The main improvements
are (i) the use of an arbitrary number of grids to describe the stellar 
interior, which allows a greater diversity of magnetic field configurations,
and (ii) the possibility of a type I superconductor interior. We
report to BBGN for details about the relativistic formulation of Maxwell
equations and the technique to solve them. Let us simply recall here
that the obtained solutions are fully relativistic and 
self-consistent, all the effects of the
electromagnetic field on the star's equilibrium (Lorentz force, spacetime
curvature generated by the electromagnetic stress-energy) being taken into
account. The magnetic field is axisymmetric and poloidal.  
The numerical technique is based on a spectral method (numerical details
can be found in Bonazzola et al. 1993). 

Thanks to the splitting 
(\ref{e:decomp,Iij}), we do not need to take into account the rotation  to 
compute the ${\cal I}_{ij}^{\rm dist}$ induced by the magnetic field. 
Consequently we consider {\em static} magnetized neutron star models. 
The reference (non-magnetized) configuration is taken to be a $1.4\, M_\odot$
static neutron star built with
the equation of state ${\rm UV}_{14}+{\rm TNI}$ 
of Wiringa, Fiks \& Fabricini (1988). This latter is a modern and 
medium stiffness equation of state (cf. Sect.~4.1.2 of Salgado et al. 1994). 
The circumferential radius is
$R=10.92\ {\rm  km}$, the baryon mass $1.56 \, M_\odot$,
the moment of inertia $I=1.23\times 10^{38} {\ \rm kg\, m}^2$
and the central value 
of $g_{00}$ is 0.36, which shows that such an object is highly relativistic. 
Various magnetic field configurations have been considered; the most 
representative of which are presented hereafter. 

\subsubsection{Normal case} \label{s:normal}

Let us first consider the case of a perfectly conducting interior (normal
matter, non-superconducting).  
As discussed in BBGN, the electric current distribution $\vec{j}$
cannot be arbitrary in order to lead to a stationary configuration: it must be
related to the covariant $\varphi$ component of the electromagnetic potential
vector, $A_\varphi$, by $j^\varphi - \Omega j^t  = (e+p) f(A_\phi)$, where
$e$ and $p$ are respectively the proper energy density and pressure of the 
fluid and $f$ is an arbitrary function. The simplest magnetic configuration 
is given by the choice $f(x) = {\rm const}$. It results in electric currents in the
whole star with a maximum value at half the stellar radius in the equatorial plane. 
The corresponding magnetic field
distribution is shown (in coordinate space)
in Fig.~\ref{f:mag:f=const}. The resulting 
distortion factor is $\beta=1.01$, which is above the $1/5$ value of the 
uniform magnetic field/incompressible fluid Newtonian model considered in 
Sect.~\ref{s:incomp}, but still very low. This is not surprising since
the magnetic field configuration has a very simple structure: it is certainly
not the configuration which maximizes the deformation at fixed magnetic dipole
moment. 

Changing the function $f$ does not lead to a dramatic increase in $\beta$:  
for $f(x) = \alpha_1 ( x \exp(-x^2) + \alpha_2)$, we get $\beta=1.07$ and
for $f(x) = \alpha_1 \cos(x/\alpha_2)$, $\beta = 1.29$. 

Our multi-grid code allows to study localized distribution of the electric
current. Figure~\ref{f:mag:j:coeur} corresponds to the electric current
distribution given by $f={\rm const}$ from $r=0$ up to $r_*=0.5 \, r_{\rm eq}$, where
$r_{\rm eq}$ is the $r$ coordinate of the equator, 
and to no electric current for $r>r_*$. The distortion factor
is only $\beta=0.70$. This can be understood since in the absence of
electric current, there is no Lorentz force in the outer part of the star. 
For electric currents concentrated deep in the stellar core, the situation is more
favorable. Indeed, since in the regions where $\vec{j}=0$ the magnetic field
$\vec{B}$  falls off as $\sim r^{-3}$, a moderate value of ${\cal M}$ can
correspond to an important value of $\vec{B}$ at the stellar centre, leading to 
a substantial deformation of the core, that gravitationally influences
the rest of the star. For instance, for 
$r_* = 0.1 \, r_{\rm eq}$, we get $\beta= 5.86$. 

\begin{figure}
\epsfig{figure=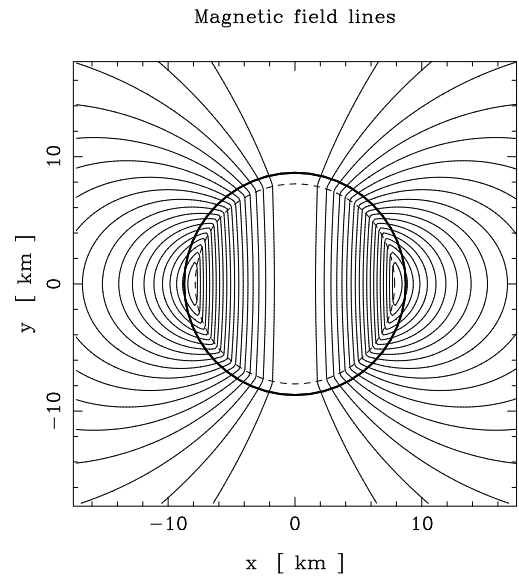,angle=270,height=8cm}
\caption[]{\label{f:mag:j:croute}
Magnetic field lines generated by a current distribution localized in 
the crust of the star and corresponding to
the choice $f={\rm const}$. The thick line denotes the star's surface
and the dashed line the internal limit of the electric current distribution.  
The distortion factor corresponding to this situation is $\beta=8.84$.}
\end{figure}

The opposite situation corresponds to electric currents localized in the neutron
star crust only. Figure~\ref{f:mag:j:croute} presents one such configuration: 
the electric current is limited to the zone $r>r_*= 0.9 \, r_{\rm eq}$
($f(x)={\rm const.}$).  
The resulting distortion factor is $\beta=8.84$.

\begin{figure}
\epsfig{figure=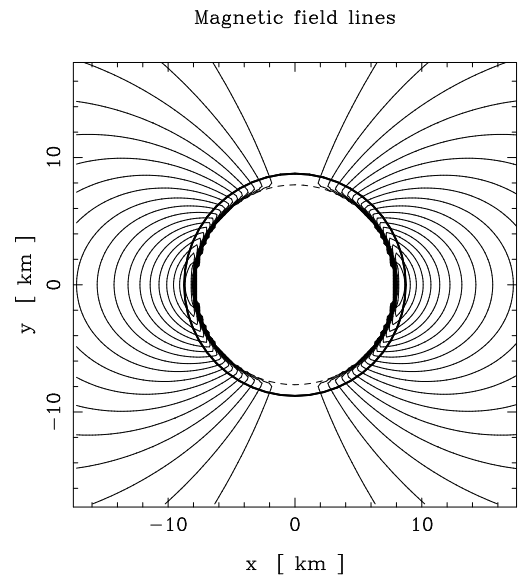,angle=270,height=8cm}
\caption[]{\label{f:mag:supra}
Magnetic field lines generated by a current distribution exterior to a 
type I superconducting core. The thick line denotes the star's surface
and the dashed line the external limit of the superconducting region. 
The distortion factor corresponding to this situation is $\beta=157$.}
\end{figure}

\begin{figure}
\epsfig{figure=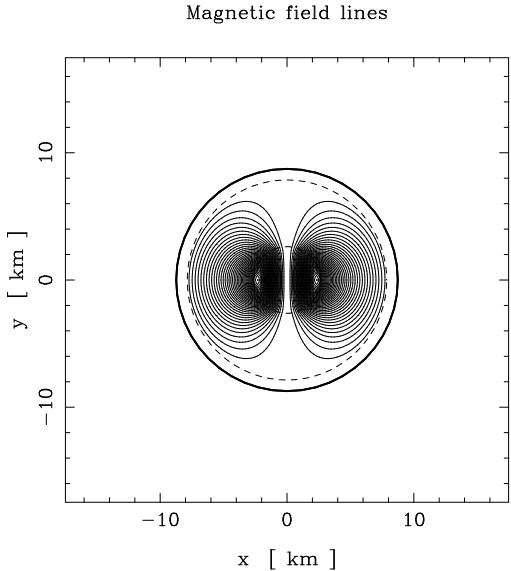,angle=270,height=8cm}
\caption[]{\label{f:mag:contra}
Magnetic field lines generated by some electric current rotating in one
direction for $0\leq r \leq 0.3\, r_{\rm eq}$ and in the opposite direction
for $0.9 \, r_{\rm eq} \leq r \leq r_{\rm eq}$. 
The thick line denotes the star's surface
and the dashed lines the limits of the electric current distributions. 
The distortion factor corresponding to this situation is
$\beta=5.7\times 10^3$.}
\end{figure}

\subsubsection{Type I superconductor} \label{s:supraI}

Let us consider the case of a superconducting interior, of type I, which 
means that all the magnetic field has been expulsed from the superconducting 
region. In the configuration depicted in Fig.~\ref{f:mag:supra}, the
neutron star interior is superconducting up to $r_*=0.9\, r_{\rm eq}$. 
For $r>r_*$, the matter is assumed to be a perfect conductor carrying an
electric current which corresponds to $f(x) = {\rm const}$. The resulting
distortion factor is much higher than in the normal case: $\beta=157$. 
For $r_*=0.95\, r_{\rm eq}$, $\beta$ is even higher: $\beta = 517$. 

\subsubsection{Counter-rotating electric currents}

The above values of $\beta$, of the order $10^2 - 10^3$, though much higher than
in the simple normal case (Sect.~\ref{s:normal}), are still too low to 
lead to an amplitude detectable by the first generation of 
interferometric detectors in the case of the Crab or Vela pulsar 
[cf. Eqs.~(\ref{e:th0,Crab}), (\ref{e:th0,Vela}) and (\ref{e:hmin})]. 
It is clear that the more disordered the magnetic field the higher $\beta$,
the extreme situation being reached by a stochastic magnetic
field: the total magnetic dipole moment $\cal M$ almost vanishes, in 
agreement with the observed small value of $\dot P$, whereas the mean value 
of $B^2$ throughout the star is huge. 
Note that, according to Thompson \& Duncan (1993), turbulent dynamo
amplification driven by convection in the
new-born neutron star may generate small scale magnetic fields as strong as
$3\times 10^{11}{\ \rm T}$ with low values of $B_{\rm dipole}$ outside the
star and hence a large $\beta$. 

In order to mimic such a stochastic magnetic field, let 
us consider the case of counter-rotating electric currents, namely 
$\vec{j}$ is given by (i)  $f(x) = f_1$ for 
$0\leq r \leq r_1^*$, where $f_1$ is a constant,
(ii) $f(x) = f_2$ for $r_2^*\leq r \leq r_{\rm eq}$, where
$f_2$ is a constant with the opposite sign than $f_1$ and (iii) no electric
current for $r_1^* < r < r_2^*$. Figure~\ref{f:mag:contra} corresponds to
the case $r_1^* = 0.3 \, r_{\rm eq}$, $r_2^* = 0.9\, r_{\rm eq}$ and
$f_1/f_2 = - 5.45$. The resulting distortion factor is
$\beta=5.7\times 10^3$. The value of magnetic dipole moment,
${\cal M} = 5.1\times 10^{27} {\ \rm A\, m}^2$, is similar to that
of the Crab pulsar (assuming $\alpha \sim 1$), but the amplitude of the
magnetic field at the star's centre, $B_{\rm c} = 5.7\times 10^{12} {\ \rm T}
= 5.7\times 10^{16} {\ \rm G}$ is $\sim 10^4$ times higher than the polar
value deduced from a simple dipole model.
Clearly for such configurations $\beta$ can be
made arbitrary large by adjusting the parameters $f_1$ and $f_2$.
 
\subsubsection{Type II superconductor}

It is not clear if the protons of the neutron star interior form a type I
(Sect.~\ref{s:supraI}) or a type II superconductor (P.~Haensel, private
communication). In the latter case, the magnetic field inside the star
is organized in an array of quantized magnetic flux tubes, each tube containing
a magnetic field $B_{\rm c} \sim 10^{11} \ {\rm T}$ (Ruderman 1991).
Besides, the neutrons constitute a superfluid, with quantized vortices.
As the neutron star spins down, the neutron vortices migrate away from
the rotation axis. As discussed by Ruderman (1991, 1994),
the magnetic flux tubes are forced to move with them. However, they
are pinned in the highly conducting crust. This results in
crustal stresses of the order $B_{\rm c} B / 2\mu_0$
[Ruderman 1991, Eq.~(10)], where $B$ is the mean value of the magnetic
field in the crust ($B\sim 10^8 \ {\rm T}$ for typical pulsars). This
means that the crust is submitted to stresses $\sim 10^3$ higher than in
the uniformly distributed magnetic field considered in
Sect.~\ref{s:normal} (compare $B_{\rm c} B/2\mu_0$ with $B^2/2\mu_0$).
The magnetic distortion factor $\beta$ should increase in the same
proportion. We have not done any numerical computation to confirm this
but plan to study type II superconducting interiors in a future paper.

\begin{figure*}
\epsfig{figure=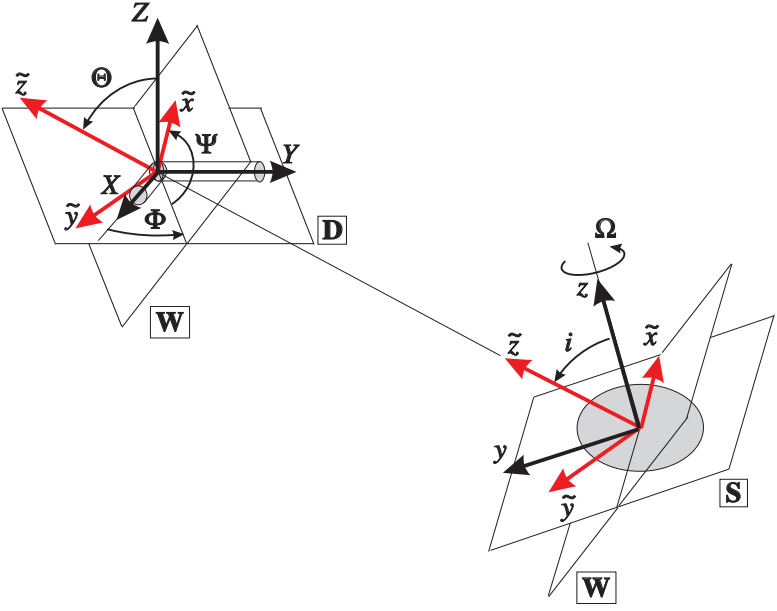,height=8cm}
\caption[]{\label{f:triades}
Relative orientation of various orthonormal frames introduced in the text:
$(\w{e}_x, \w{e}_y, \w{e}_z)$ is the neutron star inertial (non-rotating)
frame, associated with the ACMC coordinates $(x,y,z)$; $(\w{e}_x, \w{e}_y)$
define the plane {\bf S} perpendicular to the rotation axis.
$(\w{e}_{\tilde x}, \w{e}_{\tilde y}, \w{e}_{\tilde z})$ is the frame
associated with the gravitational wave received on Earth; 
$(\w{e}_{\tilde x}, \w{e}_{\tilde y})$ define the wave plane {\bf W}.
Note that $\w{e}_{\tilde x} = \w{e}_x$. 
$(\w{e}_X, \w{e}_Y,\w{e}_Z)$ is the interferometric detector frame;
$(\w{e}_X, \w{e}_Y)$ define the detector plane {\bf D}.}
\end{figure*}

\section{Signal received by an interferometric detector} \label{s:detect}

Due to the weakness of the expected gravitational signal from pulsars, 
long integration times, typically of the order of the year, are required
to extract the signal out of the noise. For such a long observing time, the
motion of the Earth must be taken into account for it modifies the position
of the pulsar with respect to the antenna pattern of the 
interferometric detector. It results in a modulation
of both the amplitude and the frequency of the signal 
(Jotania, Valluri \& Dhurandhar 1995). Note that if one is searching for a
known pulsar, the frequency modulation (Doppler shift) can be obtained readily 
from the radio observations. 

In this section, we examine the daily modulation of the signal amplitude due to 
the Earth's rotation. The special case of an interferometric detector 
situated on the equator with arms symmetrically placed about the North-South
direction and a gravitational wave coming from the Northern celestial pole
has been treated by Jotania et al. (1995). 

\subsection{Beam-pattern factors}

\begin{figure}
\epsfig{figure=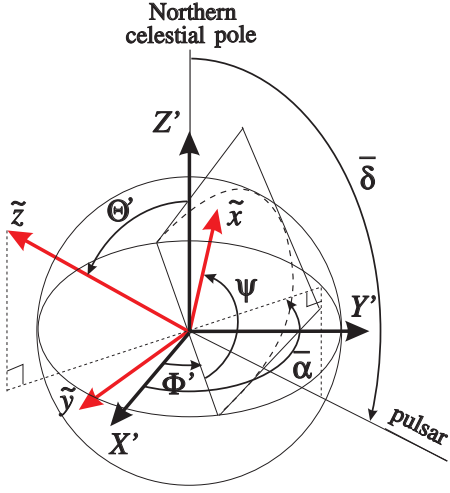,height=8cm}
\caption[]{\label{f:celeste}
Relative orientation of the gravitational wave frame 
$(\w{e}_{\tilde x}, \w{e}_{\tilde y}, \w{e}_{\tilde z})$ 
and the celestial sphere frame $(\w{e}_{X'},\w{e}_{Y'},\w{e}_{Z'})$.}
\end{figure}

Let $(\w{e}_{\tilde x}, \w{e}_{\tilde y}, \w{e}_{\tilde z})$ be the
orthonormal 
frame associated with the gravitational wave: $\w{e}_{\tilde z}$ is
perpendicular to the wave plane and parallel to the ``line of sight'',
being orientated from the neutron star centre to the Earth centre;
$\w{e}_{\tilde x}$ is the unit vector of the wave plane which is perpendicular
to the star's rotation axis. In the 
$(\w{e}_{\tilde x}, \w{e}_{\tilde y}, \w{e}_{\tilde z})$ frame
the transverse traceless gravitational wave is expressible from its two 
polarization modes $h_+$ and $h_\times$ as given by 
Eqs.~(\ref{e:h+,gen})-(\ref{e:hx,gen}) or 
(\ref{e:h+,mag})-(\ref{e:hx,mag}):
\be
    \w{h} = h_+ \, (\w{e}_{\tilde x} \otimes \w{e}_{\tilde x}
	- \w{e}_{\tilde y} \otimes \w{e}_{\tilde y} )
  	+ h_\times \, ( \w{e}_{\tilde x} \otimes \w{e}_{\tilde y}
	+ \w{e}_{\tilde y} \otimes \w{e}_{\tilde x} ) \ .
\ee
The response of an interferometric detector of the VIRGO/LIGO
type to the above gravitational wave depends upon the
relative orientation of wave frame with respect to the detector's arms.
Let $(\w{e}_X, \w{e}_Y,\w{e}_Z)$ be the orthonormal frame such that 
$\w{e}_Z$ is perpendicular to the detector plane, pointing toward the zenith
and $\w{e}_X$ and $\w{e}_Y$ are unit vectors along the two detector's arms. 
Let $(\Theta,\Phi,\Psi)$ be the three 
Euler angles which specify the position
of the wave frame $(\w{e}_{\tilde x}, \w{e}_{\tilde y}, \w{e}_{\tilde z})$ 
with respect to the detector frame $(\w{e}_X, \w{e}_Y,\w{e}_Z)$
(cf. Fig.~\ref{f:triades}). The signal measured by the detector is
\be
h(t) = F_+(\Theta,\Phi,\Psi) \, h_+(t) + 
       F_\times(\Theta,\Phi,\Psi) \, h_\times(t) \ ,
\ee
with the following beam-pattern factors
[cf. e.g. Eqs.~(103)-(104) of Thorne (1987) or Eq.~(7a) of
Dhurandhar \& Tinto (1988)] :
\begin{eqnarray}
  F_+(\Theta,\Phi,\Psi) & = & {1\ov 2} (1+\cos^2\Theta) \cos 2\Phi 
		\cos 2\Psi \nonumber \\
 & & \qquad - \cos\Theta \sin 2 \Phi \sin 2 \Psi 
					\label{e:F+} \\
  F_\times(\Theta,\Phi,\Psi) & = & {1\ov 2} (1+\cos^2\Theta) \cos 2\Phi
	\sin 2\Psi \nonumber \\
 & & \qquad + \cos\Theta \sin 2\Phi \cos 2\Psi 
					\label{e:Fx} \ .
\end{eqnarray}
The Euler angles $(\Theta,\Phi,\Psi)$ are not constant because of the
motion of the detector with respect to the source induced by the Earth's
diurnal rotation and revolution around the Sun. To compute their variations
let us introduce the ``celestial sphere frame'' $(\w{e}_{X'},\w{e}_{Y'},
\w{e}_{Z'})$ such that $\w{e}_{Z'}$ is along the Earth's rotation axis, 
pointing toward the North pole, $\w{e}_{X'}$ and $\w{e}_{Y'}$ are
in the Earth's equatorial plane, $\w{e}_{X'}$ pointing toward the vernal point
(i.e. $\w{e}_{X'}$ is along the intersection of the Earth's equatorial plane
with the Earth's orbital plane). For time scales of the order of the year, 
$(\w{e}_{X'},\w{e}_{Y'},\w{e}_{Z'})$ can be considered as fixed with 
respect to the (approximatively) inertial frame containing the
solar system barycentre and the neutron star centre. The wave frame
$(\w{e}_{\tilde x}, \w{e}_{\tilde y}, \w{e}_{\tilde z})$ is also
fixed with respect to this inertial frame. Then let $(\Theta',\Phi',
\psi)$ be the Euler angles which specify the position of the wave frame
with respect to the celestial sphere frame (cf. Fig.~\ref{f:celeste}). 
$\Theta'$ and
$\Phi'$ are simply related to the equatorial coordinates of the neutron
star on the celestial sphere (the right ascension\footnote{a bar
is put on $\alpha$ to distinguish it from the angle $\alpha$ between
the magnetic and rotation axis introduced earlier in the text.} $\balpha$ and 
the declination $\bdelta$) by (cf. Fig.~\ref{f:celeste}):
\be \label{e:Theta',Phi'}
	\Theta' = \bdelta + {\pi \ov 2} \qquad \mbox{and} \qquad
	\Phi' = \balpha - {\pi \ov 2} \ .
\ee
The triad $(\w{e}_{\tilde x}, \w{e}_{\tilde y}, \w{e}_{\tilde z})$
is related to the triad $(\w{e}_{X'},\w{e}_{Y'},\w{e}_{Z'})$ by 
\be \label{e:matA}
	\w{e}_{\tilde i} = A_{\tilde i J'} \ \w{e}_{J'} \ ,
\ee
with the orthogonal matrix  
[cf. e.g. Eq.~(4-46) of Goldstein (1980) with the substitution
(\ref{e:Theta',Phi'})] 
\begin{eqnarray}
 A_{\tilde i J'}\!\!\!& & =   \l( \begin{array}{c}
      \sin\balpha \cos\psi - \cos\balpha \sin\bdelta \sin\psi  \\
      -\sin\balpha \sin\psi - \cos\balpha \sin\bdelta \cos\psi  \\
      -\cos\balpha \cos\bdelta 
	\end{array} \r. \nonumber \\
  & & \l. \begin{array}{cc}
      -\cos\balpha \cos\psi - \sin\balpha\sin\bdelta\sin\psi &
      \cos\bdelta \sin\psi \\
       \cos\balpha \sin\psi - \sin\balpha\sin\bdelta \cos\psi &
      \cos\bdelta \cos\psi \\
      -\sin\balpha \cos\bdelta &
      -\sin\bdelta
	\end{array} \r)	. 
\end{eqnarray}

\begin{figure}
\epsfig{figure=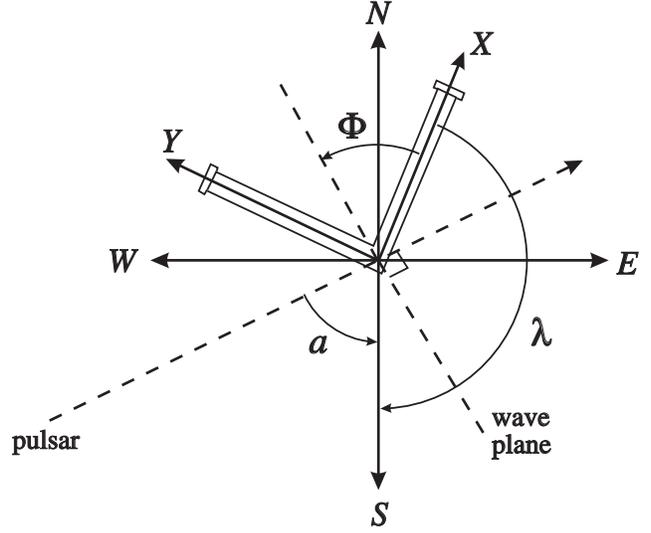,height=7cm}
\caption[]{\label{f:detecteur}
Orientation of the detector arms with respect to the local North-South
and West-East directions. The dashed line is the projection of the
gravitational wave propagation direction on the horizontal plane.
The depicted configuration corresponds to that of the VIRGO detector
($\lambda=-161^o\, 33'$).}
\end{figure}

Let $(\w{e}_{X''},\w{e}_{Y''},\w{e}_{Z''})$ be the orthonormal frame linked
to the geographical location of the detector site, such that 
$\w{e}_{Z''} = \w{e}_Z$ is the local vertical, $\w{e}_{X''}$ is in the
North-South direction and $\w{e}_{Y''}$ is in the West-East direction
(cf. Fig.~\ref{f:detecteur}). The position of the wave frame 
$(\w{e}_{\tilde x}, \w{e}_{\tilde y}, \w{e}_{\tilde z})$ with respect
to the cardinal frame $(\w{e}_{X''},\w{e}_{Y''},\w{e}_{Z''})$
is determined by the three Euler angles $(\Theta,\Phi'',\Psi)$ 
where $\Theta$ and $\Psi$ are the same angles as those relative to 
the detector frame and entering Eqs.~(\ref{e:F+})-(\ref{e:Fx}). The
angle $\Phi''$ is related to $\Phi$ by
\be \label{e:Phi-Phi''}
	\Phi = \Phi'' + \lambda \ ,
\ee
where $\lambda$ is the azimuth of the $X$-arm of the detector, i.e.
the angle between the South direction and the $X$-arm, as measured in the
retrograd way. $\Phi''$ is directly related to the azimuth $a$ of the source
by
\be \label{e:Phi''-a}
     \Phi'' = - a - {\pi \ov 2} \ .
\ee
The minus sign which occurs in this relation comes from the fact that
the azimuth is measured {\em westwards} from the South, hence in the
inverse trigronometric way. 
Putting together relations (\ref{e:Phi-Phi''}) and (\ref{e:Phi''-a})
we obtain
\be \label{e:Phi(a)}
    \Phi = - a + \lambda - {\pi \ov 2}  \ .
\ee
Let $(\theta'',\varphi'',\psi'')$ be the Euler angles which specify the 
position of the cardinal frame $(\w{e}_{X''},\w{e}_{Y''},\w{e}_{Z''})$
with respect to the celestial sphere frame
$(\w{e}_{X'},\w{e}_{Y'},\w{e}_{Z'})$. One has immediatly $\psi''=-\pi/2$
because $\w{e}_{X''}$ is orientated in the North-South direction. 
$\theta''$ is simply related to the latitude $l$ of the detector site
by
\be
	\theta'' = {\pi \ov 2} - l  \ .
\ee
$\varphi''$ is linked to the local sidereal time $T$ (i.e. the angle between
the local meridian and the vernal point) by
\be
    \varphi'' = T + {\pi \ov 2} \ .
\ee
$T$ can be expressed in terms of the sidereal time at Greenwich at
0h UT, $T_{\rm Green}(0)$, and the local UT time $t$ by
\be
   T(t) = \sigma t + T_{\rm Green}(0) - L  \ ,
\ee
where $\sigma = 1.00273790935 \times 15^o/{\rm h}$ (Meeus 1991) 
is a conversion factor from mean solar time
to sidereal time (hence $\sigma$ accounts for the revolution of the
Earth around the Sun) and $L$ is the longitude of the detector site
(the minus sign in front of $L$ comes from the fact that the
geographical longitudes are measured positively {\em westwards}). 
The triad $(\w{e}_{X''}, \w{e}_{Y''}, \w{e}_{Z''})$
is related to the triad $(\w{e}_{X'},\w{e}_{Y'},\w{e}_{Z'})$ by 
\be \label{e:matB}
	\w{e}_{I''} = B_{I''J'} \ \w{e}_{J'} \ ,
\ee
with the orthogonal matrix
\be
 B_{I''J'} =    \l( \begin{array}{ccc}
	\sin l \cos T  &  \sin l \sin T &  - \cos l \\
        -\sin T   & \cos T  & 0 \\
   	\cos l \cos T &   \cos l \sin T & \sin l  
      \end{array} \r)  \ .
\ee
Now the wave frame $(\w{e}_{\tilde x}, \w{e}_{\tilde y}, \w{e}_{\tilde z})$
is related to the triad $(\w{e}_{X''},\w{e}_{Y''},\w{e}_{Z''})$ by 

\be \label{e:matC}
   \w{e}_{\tilde i} = C_{\tilde i J''} \ \w{e}_{J''} \ ,
\ee
with the orthogonal matrix 
\begin{eqnarray}
 C_{\tilde i J''}\!\!\!& & =   \l( \begin{array}{c}
	\cos\Psi \cos\Phi'' - \cos\Theta \sin \Phi'' \sin\Psi  \\
	-\sin\Psi\cos\Phi'' - \cos\Theta\sin\Phi''\cos\Psi  \\
	\sin\Theta\sin\Phi'' 
	\end{array} \r. \nonumber \\
  & & \l. \begin{array}{cc}
	\cos\Psi \sin\Phi'' + \cos\Theta \cos\Phi'' \sin\Psi &
	\sin\Psi \sin\Theta \\
	-\sin\Psi\sin\Phi'' + \cos\Theta\cos\Phi''\cos\Psi &
	\cos\Psi\sin\Theta \\
	-\sin\Theta\cos\Phi'' &
	\cos\Theta 
	\end{array} \r)	. \label{e:C}
\end{eqnarray}
From equations~(\ref{e:matA}), (\ref{e:matB}) and (\ref{e:matC}), the matrix
$C$ is given by the product
\be
	C = A \cdot {}^t B \ .
\ee
Performing the right-hand-side product and identifying each matrix element
by those of expression (\ref{e:C}) leads to the following
trigonometrical relations
\be \label{e:cosTheta}
  \cos\Theta = - \cos\bdelta \cos l \cos H(t) - \sin \bdelta \sin l
\ee
\be \label{e:cosa}
   \sin\Theta\cos a = \cos\bdelta \sin l \cos H(t) - \sin\bdelta \cos l
\ee
\be \label{e:sina}
   \sin\Theta \sin a = \cos\bdelta \sin H(t) 
\ee
\begin{eqnarray} 
   \ \!\!\!\!\!\!& & \sin\Theta \sin \Psi  = - \cos \psi \cos l \sin H(t) 
				\nonumber \\	
	& & \qquad -  \sin \psi \sin\bdelta \cos l \cos H(t)
	+ \sin \psi  \cos\bdelta \sin l \label{e:sinPsi}
\end{eqnarray}
\begin{eqnarray}  
   \ \!\!\!\!\!\!& & \sin\Theta \cos \Psi  =  \cos\psi \cos\bdelta \sin l  
				\nonumber \\
	& & \qquad + \sin\psi \cos l \sin H(t) - \cos \psi \sin \bdelta 
		\cos l \cos H(t) , \label{e:cosPsi}
\end{eqnarray}
where we have used the relation (\ref{e:Phi''-a}) between 
$\Phi''$ and $a$ and have introduced the local hour angle of the source:
\be \label{e:H(t)}
	H(t) = T(t) - \balpha = \sigma t + T_{\rm Green}(0) - L  
				- \balpha \ .
\ee

From the above relations the beam-pattern factors $F_+(\Theta,\Phi,\Psi)$
and $F_\times(\Theta,\Phi,\Psi)$ can be computed for any instant $t$,
given the position $(l,L)$ of the detector on the Earth, its orientation
$\lambda$, and the position $(\balpha,\bdelta)$ of the source on the sky,
as well as the angle $\psi$ that the vector $\w{e}_{\tilde x}$ forms
with the intersection line of the wave plane and the Earth's equatorial plane.   
First one computes the local hour angle $H(t)$ by Eq.~(\ref{e:H(t)}), 
and the angle $\Theta$ by Eq.~(\ref{e:cosTheta}). The angle $\Phi$
is computed by means of Eqs.~(\ref{e:Phi(a)}), (\ref{e:cosa})
and (\ref{e:sina}). Finally, the local polarization angle $\Psi$ is deduced
from Eqs.~(\ref{e:sinPsi}) and (\ref{e:cosPsi}). 

\begin{figure*}
\epsfig{figure=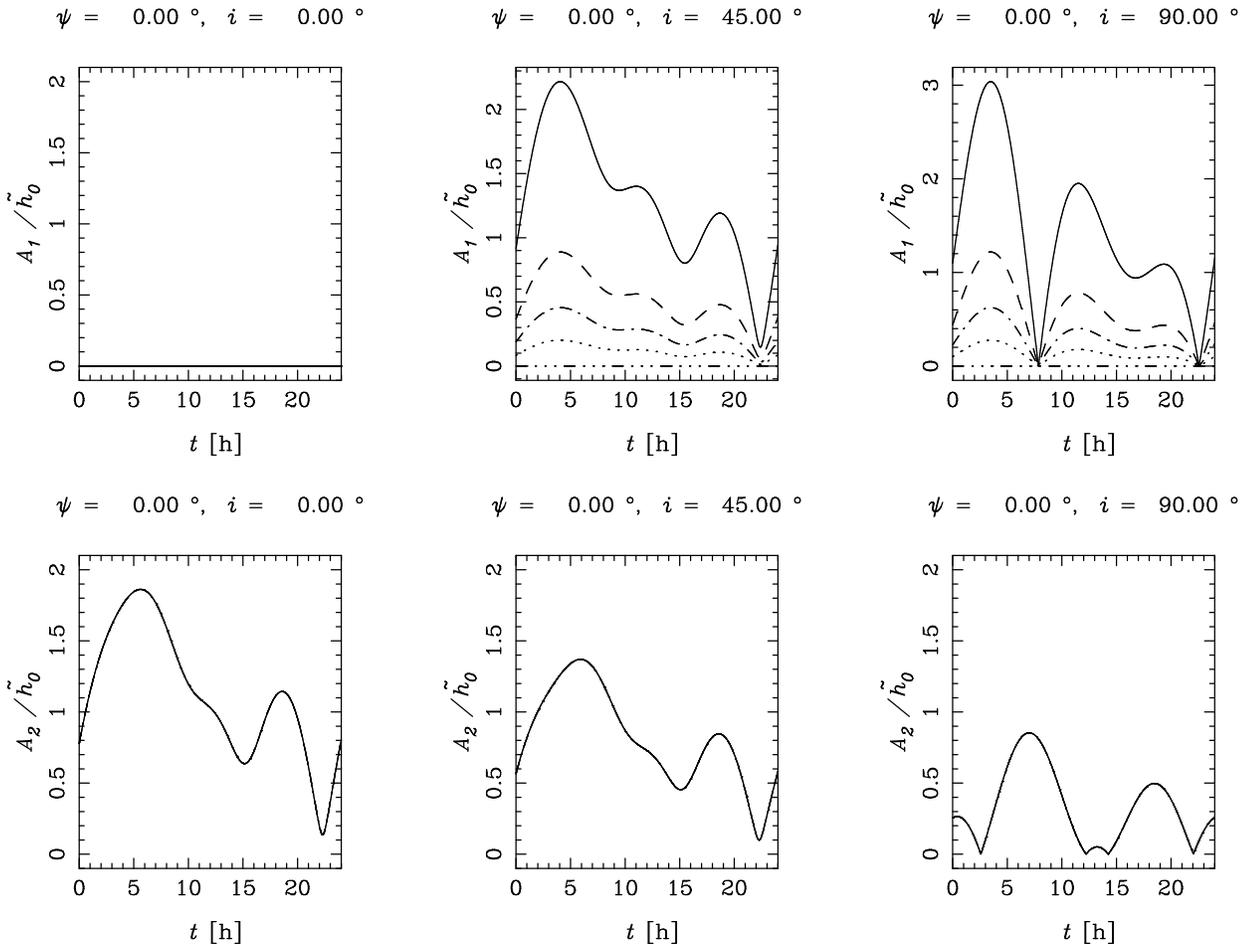,angle=270,height=13cm}
\caption[]{\label{f:amplitude,1}
Relative amplitude of the signal at the frequency $\Omega$ ($A_1$, top) and
$2\Omega$ ($A_2$, bottom) as a function of the local UT time $t$ (assuming a vanishing
local sidereal time at 0 h UT), for the specific
case of the VIRGO detector and the Crab pulsar. The polarization angle
is $\psi = 0^o$. Each plot corresponds to a given value of the inclination angle
$i$. Various lines in the same plot correspond to various values for the angle
$\alpha$ between the magnetic axis and the rotation axis: solid line: 
$\alpha=15.00^o$, dashed line: $\alpha = 33.75^o$, dot-dash-dot-dash line: 
$\alpha=52.50^o$, dotted line: $\alpha = 71.25^o$, dash-dot-dot-dot line:
$\alpha = 90.00^o$. $\tilde h_0$ is defined by Eq.~(\ref{e:def:th0}).}
\end{figure*}

\begin{figure*}
\epsfig{figure=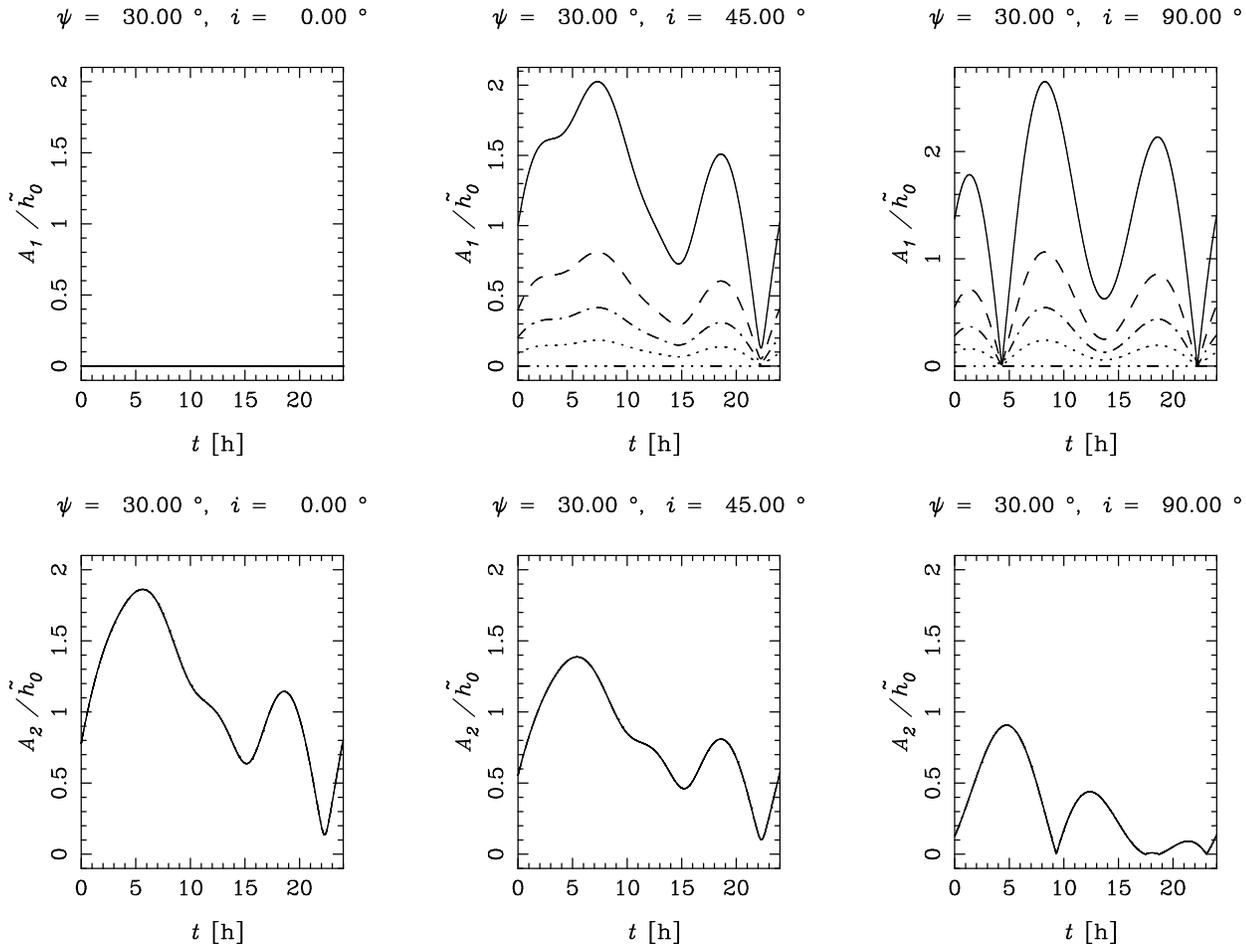,angle=270,height=13cm}
\caption[]{\label{f:amplitude,2}
Same as Fig.~\ref{f:amplitude,1} but for the polarization angle 
$\psi = 30^o$.}
\end{figure*}

\begin{figure*}
\epsfig{figure=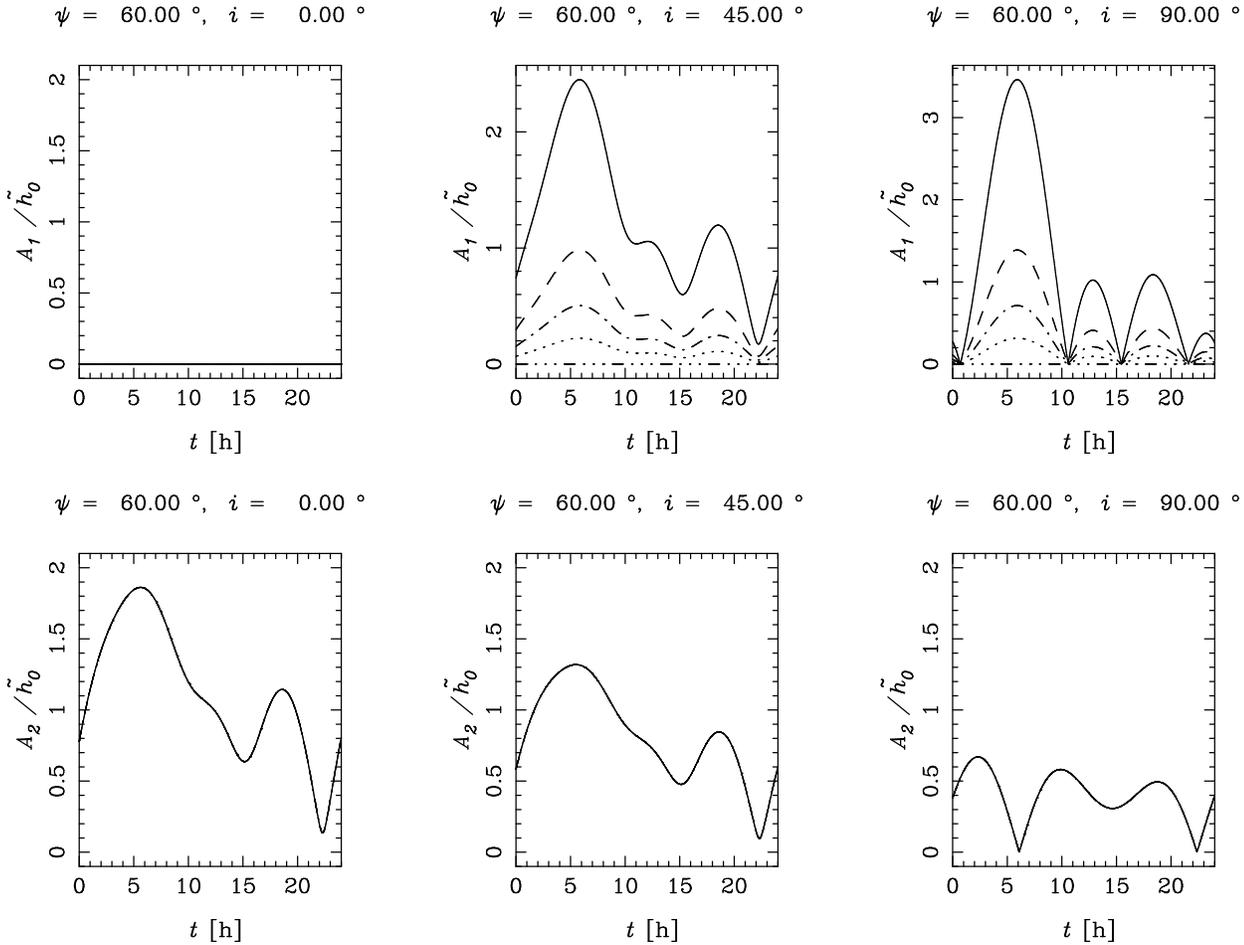,angle=270,height=13cm}
\caption[]{\label{f:amplitude,3}
Same as Fig.~\ref{f:amplitude,1} but for the polarization angle 
$\psi = 60^o$. The case $\psi = 90^o$ is identical to $\psi = 0^o$ 
(Fig.~\ref{f:amplitude,1}).}
\end{figure*}

\subsection{Signal from a rotating magnetized neutron star}

According to Eqs.~(\ref{e:h+,mag})-(\ref{e:hx,mag}), 
the gravitational wave signal from a rotating neutron star slightly deformed
by its magnetic field is
\begin{eqnarray}
h_+(t) & = & h_{+1} \cos\Omega(t-t_0) + h_{+2} \cos 2\Omega(t-t_0) 
				\label{e:h+(t)} \\
h_\times(t) & = & h_{\times 1} \sin\Omega(t-t_0) + 
	h_{\times 2} \sin 2\Omega(t-t_0) \label{e:hx(t)} \ , 
\end{eqnarray}
with 
\begin{eqnarray}
  h_{+1} & = &  \tilde h_0 {\sin i \cos i \ov 2 \tan \alpha} \\
  h_{+2} & = & - \tilde h_0 (1+\cos^2 i) / 2 \\
  h_{\times 1} & = & \tilde h_0 {\sin i \ov 2 \tan  \alpha} \\
  h_{\times 2} & = & - \tilde h_0  \cos i \\
  \tilde h_0 & := & 6 \beta {R^2\ov cr} {\dot P \ov P} \label{e:def:th0}
\end{eqnarray}
Note that in Eqs.~(\ref{e:h+(t)})-(\ref{e:hx(t)}) $t_0$ accounts for a 
different time origin between the neutron
star frame (where Eqs.~(\ref{e:h+,mag})-(\ref{e:hx,mag}) have been derived) 
and the detector frame. 
Formally $t_0$ accounts also for the propagation delay $r/c$.
 
According to the above formul\ae, 
when the location $(\balpha,\bdelta)$ of the source is known, the signal 
$h(t)$ measured by the detector depends on four a priori unknown parameters: 
\begin{itemize}
\item the inclination angle $i$ of the line of sight with respect to 
the rotation axis of the star;
\item the angle
$\alpha$ between the magnetic axis and the rotation axis;
\item the polarization angle $\psi$ (cf. Figs.~\ref{f:triades} and 
\ref{f:celeste}) :
 $\psi$ measures the orientation of the major
axis of the ellipse formed by the projection of the neutron star's equator
on the plane of the sky;
\item the time $t_0$ such that $t_0 - r/c$ is the 
instant when the magnetic dipole moment vector {\boldmath $\cal M$}
is in the plane formed by the
rotation axis and the line of sight, and when the scalar product 
$\mbox{\boldmath $\cal M$}\cdot \w{e}_{\tilde z}$ is positive. 
\end{itemize}
The signal measured by the detector can be written with the
explicit dependence upon these parameters:
\begin{eqnarray}
  h(t) & = & F_+\l( {t\ov t_{\rm SD}}, \psi \r) \, h_{+1}(i,\alpha)
	\cos\Omega(t-t_0) \nonumber \\
     & & +  F_\times\l( {t\ov t_{\rm SD}}, \psi \r) \, h_{\times 1}(i,\alpha)
	\sin\Omega(t-t_0) \nonumber \\
  & & + F_+\l( {t\ov t_{\rm SD}}, \psi \r) \, h_{+2}(i)
	\cos 2\Omega(t-t_0) \nonumber \\  
  & &   + F_\times\l( {t\ov t_{\rm SD}}, \psi \r) \, h_{\times 2}(i)
	\sin 2\Omega(t-t_0)   
\end{eqnarray}
where $t_{\rm SD}$ is the duration of one sidereal day: $F_+$ and 
$F_\times$ are periodic functions of $t/t_{\rm SD}$, with period one.  

Let us introduce the amplitudes of the signal at the frequencies
$\Omega$ and $2\Omega$ respectively:
\begin{eqnarray}
  A_1 & := & \sqrt{ F_+^2 \, h_{+1}^2 + F_\times^2 \, h_{\times 1}^2 } \\
  A_2 & := & \sqrt{ F_+^2 \, h_{+2}^2 + F_\times^2 \, h_{\times 2}^2 }
\end{eqnarray}
The daily variation of $A_1$ and $A_2$, computed from 
Eqs.~(\ref{e:F+})-(\ref{e:Fx}) and (\ref{e:cosTheta})-(\ref{e:H(t)}),
is represented in Figs.~\ref{f:amplitude,1}-\ref{f:amplitude,3}
for the specific case of the VIRGO detector [$L=-10^o\, 30'$, 
$l=+43^o\, 40'$ and $\lambda=-161^0\, 33'$ (P.~Hello, private communication)]
and the Crab pulsar ($\balpha = 5\, {\rm h} \, 34\, {\rm min}$, 
$\bdelta = +22^o\, 01'$). Note that generally the amplitude of the signal 
at both $\Omega$ and $2\Omega$ 
is maximum within a few hours of the instant when the Crab pulsar crosses
the local meridian (local sidereal time $T= 5\, {\rm h} \, 34\, {\rm min}$). 
Figs.~\ref{f:amplitude,1}-\ref{f:amplitude,3} show that the measure of the
time-varying amplitude at one of the two frequencies $\Omega$ and $2\Omega$
allows to determine the polarization angle $\psi$ and the inclination angle
$i$. If the signal is recorded at both frequencies, the angle $\alpha$
between the magnetic axis and the rotation axis can be determined as well. 
Note that this angle is a fundamental parameter for the theory of pulsar 
magnetospheres. 

\section{Conclusion}

We have considered the gravitational radiation emitted by a distorted
rotating fluid star. The distortion is supposed to be symmetric with
respect to some axis which does not coincide with the rotation axis.
The gravitational emission takes place at two frequencies: $\Omega$
and $2\Omega$, where $\Omega$ is the rotation frequency, except in the
particular case where the distortion axis is perpendicular to the
rotation axis (only the frequency $2\Omega$ is then present).
As an application, the magnetic field induced deformation is treated.
If, as usually admitted, the period derivative, $\dot P$, of pulsars is 
a measure of their magnetic dipole moment, 
the gravitational wave amplitude can be related to the observable
parameters $P$ and $\dot P$ of the pulsars and to a factor $\beta$
which measures the distortion response of the star to a given magnetic
dipole moment. $\beta$ depends on the nuclear matter equation of state and
on the magnetic field distribution. The amplitude at the frequency $2\Omega$,
expressed in terms of $P$, $\dot P$ and $\beta$, is independent of the
angle $\alpha$ between the magnetic axis and the rotation axis, whereas
at the frequency $\Omega$, the amplitude increases as $\alpha$ decreases. 

Using a numerical code generating self-consistent
models of magnetized neutron stars within general relativity, we have
computed the deformation for explicit models of the magnetic field distribution
and a realistic equation of state. It appeared that the distortion at fixed
magnetic dipole moment depends very sensitively on the magnetic configuration.  
The case of a perfect conductor interior with toroidal electric currents is the
less favorable one, even if the currents are concentrated in the crust. 
Stochastic magnetic fields (that we modeled by considering 
counter-rotating currents) enhance the deformation by 
several orders of magnitude and may lead to a detectable amplitude for 
a pulsar like the Crab. As concerns superconducting interiors --- the most
realistic configuration for neutron stars ---  we have studied numerically 
type I superconductors, with a simple magnetic structure outside the 
superconducting region. The distortion factor is then $\sim 10^2$ to $10^3$
higher than in the normal (perfect conductor) case, but still insufficient to 
lead to a positive detection by the first generation of kilometric 
interferometric detectors. We have not studied in details the type II 
superconductor but have put forward some argument which makes it a promising
candidate for gravitational wave detection. Due to the complicated microphysics
involved in type II superconductors we delay their study to a future paper. 
We also plan to study the deformation induced by a possible ferromagnetic
solid interior of neutron stars, as well as the effects of a strong 
{\em toroidal} internal magnetic field. 

Regarding the reception of gravitational waves from a pulsar by an
interferometric detector, we have computed
the amplitude modulation of the signal induced by the diurnal rotation of the
Earth. By inspecting the wave form, and assuming the position of the pulsar to
be known (the pulsar can be recognized by its period), 
one can determine the inclination angle of the line of sight with
respect to the pulsar rotation axis, as well as the orientation of the
pulsar equatorial plane. Moreoever, by comparing the wave forms at the 
frequencies $\Omega$ and $2\Omega$, the angle $\alpha$ between the rotation 
axis and the magnetic axis can be determined. 

Pulsars may be good candidates for the detection by the forthcoming VIRGO and
LIGO interferometric detectors. A frequently invoked mechanism for 
gravitational emission concerns asymmetries of the neutron star solid crust
and the resulting precession. In this article, we have examined instead
the bulk deformation of the star induced by its own magnetic field. 
For some configurations of the magnetic field (stochastic distribution, type II
superconductor), the deformation may be large enough to lead to a detectable
signal by VIRGO, with the total magnetic dipole moment (or equivalently the 
surface magnetic field) keeping its (relatively small) observed value.
The positive detection of gravitational waves from pulsars would lead to
some constraints on the internal magnetic field distribution, which would
be of great interest for the theories of pulsar magnetospheres. This would 
constitute an example of a significant contribution of gravitational
astronomy to classical astrophysics.

\begin{acknowledgements}
We warmly thank Fran\c cois Bondu and Patrice Hello for useful discussions
and for checking the
calculations presented in Sect.~\ref{s:detect}. We are also indebted to
Sreeram Valluri for his careful reading of the manuscript. 
The numerical calculations have been performed on Silicon Graphics
workstations purchased thanks to the support of the SPM department of the
CNRS and the Institut National des Sciences de l'Univers.
\end{acknowledgements}

\appendix
\section{From QI to ACMC coordinates} \label{s:appendQI}

Most studies of stationary 
rotating neutron stars make use of quasi-isotropic (QI) coordinates 
$(t,r',\theta',\varphi)$ (cf. the discussion in Sect.~2 of Bonazzola et al. 
1993). In these coordinates, the spatial components of the metric tensor have 
the following  asymptotic behavior:
\begin{eqnarray}
g_{r'r'} & = & 1 + {\alpha_1\ov r'} + {\beta_{11}\cos 2\theta' +\beta_{10}
	\ov {r'}^2} + O\l( {1\ov {r'}^3} \r) \label{e:g_r'r'}\\
g_{\theta'\theta'} & = & \l[ 1 + {\alpha_1\ov r'} + {\beta_{11}\cos 2\theta' 
	+\beta_{10}\ov {r'}^2} 
	+ O\l( {1\ov {r'}^3} \r) \r] {r'}^2\\
g_{\varphi\varphi} & = & \l[ 1 + {\alpha_3\ov r'} + {\beta_{30} \ov {r'}^2} 
	+ O\l( {1\ov {r'}^3} \r) \r] {r'}^2 \sin^2\theta' \ , 
\end{eqnarray}
where $\alpha_1$, $\alpha_3$, $\beta_{10}$, $\beta_{11}$ and $\beta_{30}$ are 
some constants. By comparison with the definition (\ref{e:def:ACMC:g_ij}), it 
appears that such coordinates are not ACMC to order 1: in order to be so, 
the $1/{r'}^2$ term in
$g_{r'r'}$ and $g_{\theta'\theta'}$ should not contain any $\cos 2\theta'$, i.e. 
$\beta_{11}$ should vanish. It can be seen easily that the following coordinate
transformation leads to an ACMC coordinate system $(t,r,\theta,\varphi)$:
\begin{eqnarray}
  r' & = & r + {\beta_{11} \cos^2\theta \ov r} \\
  \theta' & = & \theta - {\beta_{11} \cos\theta \, \sin\theta \ov r^2}  \ .
\end{eqnarray}
By computing the $g_{00}$ component in the $(t,r,\theta,\varphi)$ coordinates
from the components $g_{\alpha'\beta'}$ in the $(t,r',\theta',\varphi)$ 
coordinates and by identification with Eq.~(\ref{e:def:ACMC:g_00}), one obtains 
the following value of Thorne's mass quadrupole moment:
\begin{eqnarray}
   {\cal I}_{xx} & = & {\cal I}_{yy} = - {1\ov 2} {\cal I}_{zz} \\
 {\cal I}_{zz} & = & {4\ov 9} \l( \gamma_2 - M \beta_{11} \r) \label{e:IzzQI}\\
   {\cal I}_{ij} & = & 0 \quad \mbox{if}\quad i\not = j \ ,
\end{eqnarray}
where (i) $\gamma_2$ is the coefficient of $\cos2\theta'$ in the $1/ {r'}^3$ 
term of the $1/r'$ expansion of the metric component $g_{00}$ in the QI 
coordinates 
$(t,r',\theta',\varphi)$ and (ii) $M$ is half the coefficient of $1/r'$ in 
the same expansion ($M$ is nothing else than the total gravitational mass of 
the star). 

To summarize, Thorne's quadrupole moment component ${\cal I}_{zz}$ can be 
computed
via equation (\ref{e:IzzQI}) by reading off the coefficients $\beta_{11}$
and $\gamma_2$ in the $1/r'$ expansions of the metric components in the QI 
coordinates.

\end{document}